\documentclass[twocolumn]{aastex631}

\newcommand{\rrp}[1]{{{#1}}}
\newcommand{\rrptwo}[1]{{{#1}}}

\def\f1{f_{\rm I}}

\def\beq{\begin{equation}}
\def\eeq{\end{equation}}

\def\t2{\tau_{\rm II}}

\def\sigmas0{\Sigma_{\rm s,0}}

\def\apjl{ApJ}                
\def\apjs{ApJS}               
\def\ao{Appl.Optics}          
\def\aap{A\&A}                

\usepackage{inputenc}
\usepackage{mhchem}
\usepackage{mathtools}

\makeatletter
\newcommand*{\addFileDependency}[1]{
  \typeout{(#1)}
  \@addtofilelist{#1}
  \IfFileExists{#1}{}{\typeout{No file #1.}}
}
\makeatother

\shorttitle{From atmospheric composition to planet formation?}
\shortauthors{Molli\`ere et al.}
\graphicspath{{./}{figures/}}

\begin{document}

\title{Interpreting the atmospheric composition of exoplanets: \\
sensitivity to planet formation assumptions}

\author[0000-0003-4096-7067]{Paul Molli\`ere}
\affiliation{Max-Planck-Institut f\"ur Astronomie, K\"onigstuhl 17, 69117 Heidelberg, Germany}

\author[0000-0003-0448-6354]{Tamara~Molyarova}
\affiliation{Institute of Astronomy, Russian Academy of Sciences, 48 Pyatnitskaya St., Moscow, 119017, Russia}

\author[0000-0002-8868-7649]{Bertram~Bitsch}
\affiliation{Max-Planck-Institut f\"ur Astronomie, K\"onigstuhl 17, 69117 Heidelberg, Germany}

\author[0000-0002-1493-300X]{Thomas~Henning}
\affiliation{Max-Planck-Institut f\"ur Astronomie, K\"onigstuhl 17, 69117 Heidelberg, Germany}

\author[0000-0002-1448-0303]{Aaron~Schneider}
\affiliation{Centre for ExoLife Sciences, Niels Bohr Institute, {\O}ster Voldgade 5, 1350 Copenhagen, Denmark}
\affiliation{Institute of Astronomy, KU Leuven, Celestijnenlaan 200D, 3001, Leuven, Belgium}

\author[0000-0003-0514-1147]{Laura~Kreidberg}
\affiliation{Max-Planck-Institut f\"ur Astronomie, K\"onigstuhl 17, 69117 Heidelberg, Germany}

\author[0000-0002-8743-1318]{Christian~Eistrup}
\affiliation{Max-Planck-Institut f\"ur Astronomie, K\"onigstuhl 17, 69117 Heidelberg, Germany}

\author[0000-0002-9020-7309]{Remo~Burn}
\affiliation{Max-Planck-Institut f\"ur Astronomie, K\"onigstuhl 17, 69117 Heidelberg, Germany}

\author[0000-0002-9792-3121]{Evert~Nasedkin}
\affiliation{Max-Planck-Institut f\"ur Astronomie, K\"onigstuhl 17, 69117 Heidelberg, Germany}

\author[0000-0002-3913-7114]{Dmitry~Semenov}
\affiliation{Max-Planck-Institut f\"ur Astronomie, K\"onigstuhl 17, 69117 Heidelberg, Germany}
\affiliation{Department of Chemistry, Ludwig-Maximilians-Universit\"at, Butenandtstra{\ss}e 5-13, 81377 Munich, Germany}

\author[0000-0002-1013-2811]{Christoph~Mordasini}
\affiliation{Center for Space and Habitability, Universit\"at Bern, Gesellschaftsstrasse 6, 3012 Bern, Switzerland}

\author[0000-0001-8355-2107]{Martin~Schlecker}
\affiliation{Max-Planck-Institut f\"ur Astronomie, K\"onigstuhl 17, 69117 Heidelberg, Germany}

\author[0000-0002-6429-9457]{Kamber~R.~Schwarz}
\affiliation{Max-Planck-Institut f\"ur Astronomie, K\"onigstuhl 17, 69117 Heidelberg, Germany}

\author[0000-0002-6948-0263]{Sylvestre~Lacour}
\affiliation{LESIA, Observatoire de Paris, Universit\'e PSL, CNRS, Sorbonne Universit\'e, Univ. Paris Diderot, Sorbonne Paris Cité, 5 place Jules Janssen, 92195 Meudon, France}
\affiliation{Max Planck Institute for extraterrestrial Physics, Giessenbachstra{\ss}e 1, 85748 Garching, Germany}
\affiliation{European Southern Observatory, Karl-Schwarzschild-Stra{\ss}e 2, 85748 Garching, Germany}

\author{Mathias~Nowak}
\affiliation{Institute of Astronomy, University of Cambridge, Madingley Road, Cambridge CB3 0HA, United Kingdom}
\affiliation{Kavli Institute for Cosmology, University of Cambridge, Madingley Road, Cambridge CB3 0HA, United Kingdom}

\author{Matth\"aus~Schulik}
\affiliation{Imperial College London, Blackett Laboratory, Prince Consort Road, London SW7 2AZ, United Kingdom}



\begin{abstract}

Constraining planet formation based on the atmospheric composition of exoplanets is a fundamental goal of the exoplanet community. Existing studies commonly try to constrain atmospheric abundances, or to analyze what abundance patterns a given description of planet formation predicts. However, there is also a pressing need to develop methodologies that investigate how to transform atmospheric compositions into planetary formation inferences. In this study we summarize the complexities and uncertainties of state-of-the-art planet formation models and how they influence planetary atmospheric compositions. We introduce a methodology that explores the effect of different formation model assumptions when interpreting atmospheric compositions. We apply this framework to the directly imaged planet HR~8799e. Based on its atmospheric composition, this planet may have migrated significantly during its formation. We show that including the chemical evolution of the protoplanetary disk leads to a reduced need for migration. \rrp{Moreover, we find that pebble accretion can reproduce the planet's composition, but some of our tested setups lead to too low atmospheric metallicities, even when considering that evaporating pebbles may enrich the disk gas.} We conclude that the definitive inversion from atmospheric abundances to planet formation for a given planet may be \rrp{challenging}, but a qualitative understanding of the effects of different formation models is possible, \rrp{opening up pathways for new investigations.}

\end{abstract}

\keywords{Exoplanet formation (492), Exoplanet atmospheric composition  (2021)}



\section{Introduction}
\label{sect:introduction}

The distribution of bulk planetary properties such as mass, radius, and orbital parameters encode critical information that constrains planet formation models \citep[see e.g.,][]{idalin2004,alibertmordasini2005,mordasinialibert2009,hasegawapudritz2011,lambrechtsjohansen2012,bitschlambrechts2015,nayakshinfletcher2015,cridlandpudritz2016,emsenhubermordasini2020,schleckerpham2021}. In addition, the chemical composition of planet atmospheres has long been regarded as a key to unlock the process of planet formation \citep[e.g.,][]{gautierowen1989,owenencrenaz2003} and is the explicit goal of many atmospheric characterization studies \citep[see, e.g.,][for a recent review]{madhusudhan2019}.

This is because the chemical abundances of planetary atmospheres are highly complementary to bulk planetary parameters: they likely relate to the composition of the planetary building blocks in the protoplanetary disk, be it planetesimals, pebbles, or gas. The composition of the building blocks is determined by disk processes, while their relative importance and accretion location for a given planet is determined by the process of planet formation. Consequently, there have been a number of studies which investigate how planet formation may set the composition of an exoplanet, focusing on the planetary carbon-to-oxygen number ratio (C/O), nitrogen content, content in refractory material\footnote{Refractories generally encompass all those chemical species with a high condensation temperature, such that they are found in the solid phase of the protoplanetary disk, except for at the smallest orbital separations.}, or just overall metal content \citep[e.g.,][]{oeberg2011,fortneymordasini2013,marboeuf2014,madhusudhan2014,cridlandpudritz2016,mordasinivanboekel2016,madhusudhanbitsch2017,lothringerrustamkulov2020,schneiderbitsch2021,khorshidmin2021}.

 Combining observatories such as the {\it Hubble Space Telescope} ({\it HST}) and {\it Spitzer Space Telescope} allowed for a first look at atmospheric C/O values, albeit with large uncertainties \citep[e.g.,][]{lineknutson2014,benneke2015,brewerfischer2017}, or leading to controversial findings, as in the case of WASP-12b, which was claimed to be either carbon or oxygen-rich, with finally a firm water detection in transit pointing towards ${\rm C/O}\lesssim 1$ \citep[][]{madhusudhan2011,crossfield2012,swain2013,lineknutson2014,stevenson2014,benneke2015,kreidbergline2015}. Studying the bulk atmospheric enrichment of exoplanets, mostly based on water detections in {\it HST WFC3} spectra, has also been attempted, but the community would clearly benefit from data with higher signal-to-noise and larger spectral coverage to improve abundance constraints, break degeneracies with clouds, and probe additional atmospheric absorbers \citep[e.g.,][]{kreidbergbeanw43_2014,fisherheng2018,wakefordsing2018,pinhasmadhusudhan2019,welbanksmadhusudhan2019}. We note here, and discuss later, that a connection between atmosphere and bulk planet composition is far from trivial \citep[also see, e.g., the recent discussion in][and the references therein]{helledwerner2021}.

\rrp{Luckily, the quality of observational constraints on the planetary composition is expected to be rapidly improving. The advent of retrieval methods for medium- and high-resolution observations \citep[e.g.,][]{brogiline2016,brogiline2018,gibsonmerritt2020}, may allow to constrain the atmospheric volatile and refractory content \citep{lothringerrustamkulov2020} from the ground, and to trace even isotopologues \citep{mollieresnellen2019}. For example, \citet{gandhimadhusudhan2019b,pelletierbenneke2021,linebrogi2021} used high-resolution retrievals to constrain planetary C/O values, while the medium resolution retrievals of \citet{zhangsnellen2021} indeed revealed isotpologues for the first time.} Moreover, recent observations of the {\it GRAVITY} instrument at the {\it Very Large Telescope Interferometer} ({\it VLTI}), have led to some of the most precise constraints on C/O for planetary-mass objects to date \citep[][]{nowaklacour2020,mollierestolker2020}. Most importantly, the recent launch of the {\it James Webb Space Telescope} ({\it JWST}) and, later in the decade {\it ARIEL}, are expected to lead to excellent constraints on the C/O ratio, especially for transiting planets, and may probe the nitrogen and refractory content of cool planets \citep[e.g.,][]{greeneline2016,wangmiguel2017,danielskibaudino2018,tinettidrossart2018}. With these next-generation telescopes the focus will likely shift from observational uncertainties to uncertainties in the models for atmospheric characterization \citep[e.g.,][]{lineparmentier2016,fengline2016,blecicdobbsdixon2017}. The development of new characterization techniques is therefore necessary for interpreting future observations. This work has already begun \citep[e.g.,][]{caldas2019,taylor2020,feng2020,macdonald2020,pluriel2020,lacy2020,changeat2021,macdonaldlewis2021,nixonmadhusudhan2022}. 

Now that the atmospheric abundance constraints may become more precise than ever before, it is timely to revisit the justification stated for many observational campaigns: how can a planet's formation history {\it actually} be constrained, given its atmospheric abundances, and how well? What are the actual formation quantities that are constrainable? What are the major obstacles that would need to be overcome in case this is not possible? And, lastly, if such an inversion process is challenging for a single planet, could the distribution of abundance patterns be used to constrain some aspects of planet formation?

Our study aims at addressing some of the questions stated above. Specifically, we discuss planet formation and its complexities in the context of the inversion challenge (Section \ref{sect:planet_form_complexity}). In Section \ref{sect:formation_inversion}, we present a methodology that may prove useful for assessing the consequences of a given formation model choice, where we use a nested sampling method to constrain formation parameters, given the atmospheric composition, for different model assumptions. We show example applications, namely how chemical disk evolution, \rrp{or pebble drift, evaporation, and accretion}, may affect the inferred formation and migration history of the planet HR~8799e. Our method can be used to qualitatively understand differences between planet formation implementations. In Section \ref{sect:obs_future} we summarize which molecular and atomic species can and will be probed by atmospheric observations, and how these may serve to broadly inform the process of planet formation. We end with a short discussion and summary of our study in Section \ref{sect:summary}.
\section{The complexity of the planet formation problem}
\label{sect:planet_form_complexity}

\begin{figure}[t!]
\includegraphics[width=0.47\textwidth]{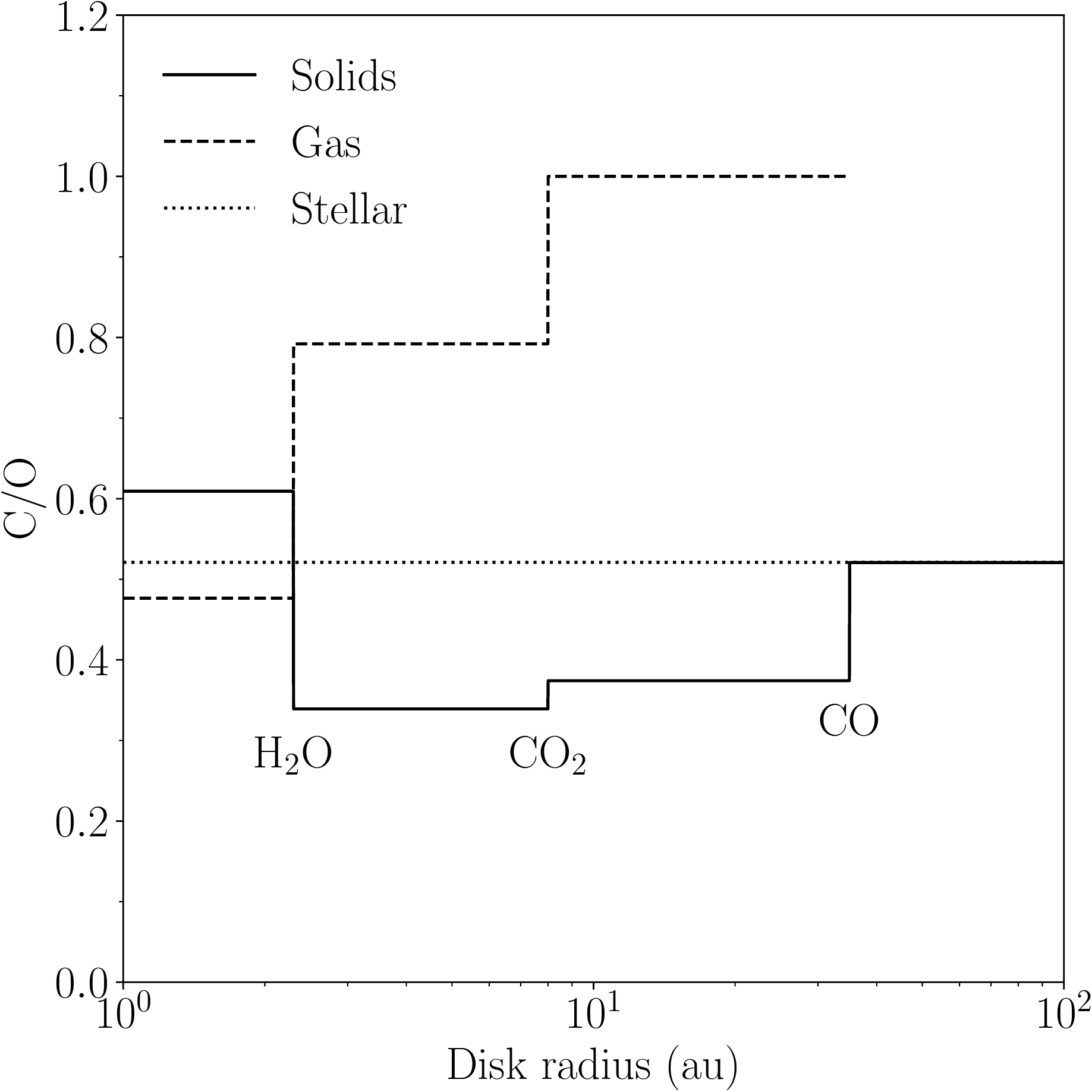}
\caption{C/O distribution of the solid and gas component in a smooth, static, circum-stellar disk assuming equilibrium condensation. The locations of the \ce{H2O}, \ce{CO2} and CO icelines are indicated in the plot.}
\label{fig:oeberg_steps}
\end{figure}

The idea of using planetary composition to constrain planet formation gained traction in the field of exoplanets with the seminal paper by \citet{oeberg2011}. Here, the authors propose that the C/O value derived from a planet's composition could be used to constrain where in a protoplanetary disk it formed. The general idea is outlined in Figure \ref{fig:oeberg_steps}, very similar to the original Figure~1 in \citet{oeberg2011}. Assuming a smooth, static, 1-dimensional disk, the authors calculated where important volatile gases such as \ce{H2O}, \ce{CO2} and CO (sorted by decreasing condensation temperature) freeze out, if present. Because for the temperature gradient in a protoplanetary disk it holds that $dT/dr<0$, where $r$ is the distance from the star, \ce{H2O} freezes out  first when moving outward, followed by \ce{CO2} and CO. This directly affects the C/O values in the gas and solid phases because water, for example, removes oxygen from the gas phase, when condensing.\footnote{We note that we assumed that 33~\% of all O and 38\% of all C is contained in the refractory solids in the example setup shown in Figure~\ref{fig:oeberg_steps}, which leads to a high solid-phase C/O inside the \ce{H2O} iceline.}

The idea then is that if the planetary C/O and overall metallicity (here: C and O content) are known, it is possible to determine where in the disk a planet has formed. This method of determining the process of planet formation has since been cited in virtually every study that aims at constraining the atmospheric composition of an exoplanet. This also has to do with the comparative ease with which the atmospheric C/O value may be constrained, as we discuss in Section \ref{sect:obs_future}. The general idea presented in \citet{oeberg2011} is powerful, but it is undeniably so that planet formation is more complicated than assumed in their study. In the following we give a summary of processes that may have to be taken into account, and inverted, when trying to connect planet composition to formation in practice.

\subsection{Disk elemental composition and structure}
\label{subsect:disk_elemental_composition_and_structure}

Constraining a planet's formation location based on its elemental abundance ratios (e.g., C/O) requires that the elemental composition of the protoplanetary disk is known. A good starting point may be to assume that the protoplanetary disk has an elemental composition that is identical to that of the host star \citep[planet formation may deplete stellar photospheres in metals with respect to the disk, however, see][]{chambers2010,bitschforsberg2018,boothowen2020}. It is therefore crucial to have knowledge about the host star's abundances that is as complete as possible, in excess of just [Fe/H], or to at least use existing scaling relations to approximate the stellar metal content for elements other than iron \citep[e.g.,][]{bitschbattistini2020}. The disk composition and assumed mass also sets an upper limit on the amount of metals that a planet may accrete during its formation \citep[e.g.,][]{baraffechabrier2008}. Alternatively, the retrieved metal content of an exoplanet may also be used to place a lower limit on the disk metal content, or even total mass, analogous to the concept of the minimum mass solar nebula \citep[e.g.,][]{hayashi1981}.

Moreover, the disk's physical and thermal structure is important to set the radial fractionation of elements into different molecular species, in solid and gaseous form, that can be accreted by a forming planet. The disk structure depends on the assumed (dust) opacities, and therefore the dust evolution \citep{schmitthenning1997,birnstielfang2016,savvidoubitsch2020}. Moreover, disks are viscously \citep{lyndenbellpringle1974} evolving \citep[or due to photoevaporation and disk winds, see][]{calrkegendrin2001,suzukiogihara2016,baiye2016,chambers2019}, prone to different instabilities \citep[e.g.,][]{flocknelson2017,klahrpfeil2018}, and will be affected by the presence of (especially massive) planets that may induce spiral density perturbations, lead to the formation of vortices, or open gaps \citep[e.g.,][]{linpapaloizou1986,cridamorbidelli2006,pinillaovelar2015,lobogomesklahr2015,binkertszulagyi2021}.

Another important effect determining the disk gas composition is the evaporation of pebbles inside of icelines, which may significantly increase the local volatile content of gas in the disk \citep{pisooeberg2015,boothilee2019,schneiderbitsch2021}. This could lead to planets being enriched in volatiles much more than expected from pure gas accretion in the classical \citet{oeberg2011} setup. \rrp{It is important to mention that the dynamics of pebbles is likewise determined by the disk structure, which sets the pebbles' growth rates, Stokes parameters (i.e., drift speeds), and may trap them into local pressure maxima \citep[e.g.,][]{paardekoopermellema2006,ormelklahr2010,birnstielklahr2012,lambrechtsjohansen2014}}.

\subsection{Disk chemistry}
\label{subsect:disk_chemistry}

The chemical composition of the protoplanetary disk \citep[e.g.,][]{henningsemenov2013} is of central importance for determining the composition of planetary building blocks, that is, the disk's gas and solid phases. In \citet{oeberg2011}, and the more recent \citet{oebergwordsworth2019} study, a simplified and static disk chemical model is assumed. In practice, the disk's chemical composition will evolve because both the disk gas and the volatile ices on grain surfaces will undergo chemical processing \citep[e.g.,][]{2016A&A...595A..83E,molyarovaakimkin2017,2018A&A...613A..14E}. This means that chemical reactions between atoms and molecules in both gas and ice may alter which molecular species are the dominant carriers of elements such as C, O, and N, over time. This is an important effect, and is expected to alter the inferred formation history of a planet, as has been pointed out by \citet{2016A&A...595A..83E}. Examples are the conversion of CO into CO$_2$ ice over time, or the conversion of \ce{N2} gas into \ce{NH3} ice \citep{molyarovaakimkin2017,2018A&A...613A..14E,2011ApJS..196...25S}. Another example of the importance of disk chemistry on planet formation are the processes that lead to the observed carbon depletion in the inner solar system \citep[e.g.,][]{mordasinivanboekel2016,cridlandeistrup2019}, \rrp{which may be attributable to the irreversible chemical destruction of carbon grains within a disk's so-called soot line, or connected to chondrule formation \citep[e.g.,][]{kresstielens2010,gailtrieloff2017,vanthoffbergin2020,libergin2021}}.

The disk chemistry itself is sensitive to many processes, such as stellar evolution \citep{2021MNRAS.500.4658M}, or the cosmic-ray ionization rate \citep{2016A&A...595A..83E,schwarzbergin2019}. Moreover, whether or not the initial composition of the disk matter is molecular, that is, 'inherited' from the composition of the natal molecular cloud, or elemental ('reset' scenario) can strongly influence the C/O values \citep{2016A&A...595A..83E}. \rrp{The disk's physical structure may also have an impact on its chemical evolution. As an example we highlight} the effect of the self-shadowing of the disk, allowing for nominally too cool compositions to occur closer to the star than otherwise expected \citep{ohnoueda2021}. 

\subsection{Planet formation}
\label{subsect:planet_formation}

\rrp{The idea that planet formation may be constrained through planet composition, as presented in \citet[][]{oeberg2011}, conceptually boils down to comparing the planetary C/O value and total metal enrichment to the C/O of the disks' solid and gaseous phases as a function of orbital distance from the star. However, } planet formation involves and connects many complex processes. This means that a forming planet cannot accrete arbitrary amounts of gas or solids at (or from) arbitrary positions in the disk.

\rrp{For example, a limiting factor for planets forming via the core accretion paradigm \citep[e.g.,][]{mizuno1980,pollackhubickyi1996}, specifically when accreting pebbles, is the pebble isolation mass.} Pebble accretion, which is the accretion of roughly cm-sized solids by the forming planet \citep[e.g.][]{ormelklahr2010,lambrechtsjohansen2012}, can only dominate the solid accretion process until the growing planet reaches this isolation mass $M_{\rm iso}$ \citep[e.g.,][]{lambrechtsjohansen2014,bitschmorbidelli2018,ataieebaruteau2018}, after which pebble accretion stops. This is because the planet induces the formation of a pressure bump in the disk, exterior to its orbit, which traps the inward-drifting pebbles. The isolation mass places an upper limit on the refractory content of a planet. A refractory content higher than allowed by this concept of $M_{\rm iso}$ could point to the importance of accreting planetesimals \citep[e.g.,][]{mordasinivanboekel2016,bruggerburn2020}, unless the planet formed very close to its star, within the refractories' icelines \citep{schneiderbitsch2021b}.

For planets growing in-situ, a planetesimal isolation mass was established by \citet{lissauerstewart1993}. It is caused by the planet depleting the local reservoir of planetesimals within the zone of its gravitational influence (the so-called `feeding-zone'). In contrast to the pebble isolation mass, the planetesimal feeding zone increases with planet mass. The different ways of how a planet's refractory content and total mass scale in planetesimal and pebble accretion models might thus allow to put limits on the contributions of the different solid reservoirs for a given planet. The planetesimal isolation mass can also be overcome via (giant) impacts or if a proto-planet migrates into regions of the disk still containing planetesimals \citep{alibertmordasini2005}.

Moreover, the accretion of gas by the growing planet is a 3-dimensional process \citep[e.g.,][]{dangelokley2003,ayliffebate2009,szulagyimorbidelli2014,ormelshi2015,schulikjohansen2019,schulikjohansen2020}. This may be especially important as the gas composition and C/O value is thought to be changing above the mid-plane of the disk \citep[e.g.,][]{molyarovaakimkin2017,cridlandbosman2020}. \rrp{Similar to the solid accretion processes, the amount of gas a planet can accrete during formation is limited. In contrast to the solids, however, the ultimately limiting factor is the lifetime of the protoplanetary disk. Once a planet enters runaway accretion, it will accrete gas as quickly as can be provided by the viscously evolving disk, until the disk dissipates. More specifically, it has been shown that gas delivery to the planet can be severely limited by gap formation, which in turn is controlled by the discs viscous resupply of the planetary feeding zone \citep[e.g.,][]{lubowseibert1999,lissauerhubickyj2009,ayliffebate2009,schulikjohansen2020,bergezcasalou2020}. For embedded planets below the gap opening mass, gas accretion may be limited if radiative cooling is counteracted by hot inflowing gas from the ambient disk. In this case even gas that enters the planetary Hill sphere is not accreted \citep[so-called `recycling',][]{cimermankuiper2017}.}

Another important process of planet formation is migration. The migration history of a planet is critical to its final composition, because it determines where in the disk it accretes material. Because the \citet{oeberg2011} approach strives to ultimately constrain formation locations with respect to the disk icelines, migration may be less of a problem if a forming planet did not migrate across icelines. Interestingly, planets forming in the inner disk may actually be trapped at locations just beyond the water iceline \citep[e.g.,][]{bitschjohansen2016,cridlandpudritz2016,muellersavvidou2021}. There also exist other traps not connected to icelines, such as the disk's dead zone inner edge \citep[][]{bitschmorbidelli2014,cridlandpudritz2016}. Qualitatively speaking, and when not currently trapped, planets are expected to migrate either by fast type-I migration or via slower type-II migration, the latter of which ensues once the planet is massive enough to open a gap in the disk \citep[see, e.g., the review by][]{baruteaubai2016}. Quantitatively, there is an ongoing debate about the actual magnitude of the torques (and therefore speed of migration), for example for type II migration \citep[e.g.,][]{duerrmannkley2015,robertcrida2018}.

Further complicating the picture are N-body interactions between the forming planets. This process is now regularly included in models of planet formation and population syntheses, but comes at an increased numerical cost \citep[e.g.,][]{alibertcarron2013,chambers2016,lambrechtsmorbidelli2019,izidorobitsch2019,emsenhubermordasini2020}. While increasing the complexity of describing planet formation,  N-body interactions may represent an alternative avenue for producing hot Jupiters, which may have been scattered in by planets further out \citep[e.g.,][]{jurictremaine2008,bitschtrifonov2020}. N-body interactions can also alter the composition of a planet via giant impacts. The amount of solids brought into a giant planet via such impacts might be substantial or even dominant compared to the amount accreted from small bodies like planetesimals or pebbles \citep[e.g.,][]{emsenhubermordasini2021b,ginzburgchiang2020,ogiharahori2021}. If the impactors originally formed in clearly different regions of the disk than the forming planet, this would blur the meaning of a well-defined formation location of a planet. Accounting for the effects of N-body interactions when trying to invert planet formation thus seems challenging. Interestingly, it has recently been shown that while N-body interactions tend to randomize the process of planet formation, machine learning techniques such as random forests still allow to predict the outcome of planet formation quite accurately \citep{schleckerpham2021}. In this work the authors show that the initial parameters of the planet formation model described in \citet{emsenhubermordasini2020}, mainly the initial location of the planetary embryo and the dust mass of the disk, may be used to predict which class a forming planet will belong to (super-Earths, Neptune-like, giant planets, etc.). These classes also correspond to certain orbital distances, and planetary compositions.

The discussion here focused mostly on planet formation via the core accretion paradigm. Other aspects are of importance if a planet forms via gravitational instability \citep[GI, see][]{boss1997}. The disk structure (and thus formation environment of the planet) will be quite different in this case. This is because GI planets may form early, when the disk is still massive, in the outer parts of the disk \citep{boss2021,schibmordasini2021}. We note that while GI is classically regarded as a way to produce wide-separation gas giant planets, it has also been suggested to allow for the formation of less massive, small-separation planets \citep{nayakshin2010}, caused by fast inward migration after formation and the associated mass loss, dubbed `tidal downsizing'.

\subsection{Planetary bulk -- atmosphere coupling}
\label{sect:planetary_bulk_atmospheric_coupling}

When aiming to constrain planet formation based on the results of atmospheric abundance characterizations one must make assumptions on how the atmospheric composition relates to the bulk composition of a planet. It has been pointed out that planets growing via core accretion may have a layer of heavily metal-enriched gas above their solid cores, due to the evaporation of pebbles and planetesimals that are destroyed when entering the hot planet's proto-atmosphere \citep[e.g.,][]{mordasinivanboekel2016,helledstevenson2017,brouwersvazan2018,brouwersormel2020}. The recent constraints by the Juno spacecraft on the interior of Jupiter are consistent with this assessment, pointing to the existence of a dilute core \citep{wahlhubbard2017,debraschabrier2019}. \rrp{We note that the exact distribution of metals in Jupiter' interior is difficult to explain, however, and may involve a formation process stretching over 2 Myr \citep[see][for a recent review]{helledstevenson2022}}. \rrp{What is more, the planet bulk metallicity constraints for transiting planets derived in \citet[][]{millerfortney2011,thorngrenfortney2016,thorngrenfortney2019b} tend to be higher than the metallicities reported for planetary atmospheres \citep[albeit with large uncertainties, see][]{welbanksmadhusudhan2019}, making an enrichment of the interior with respect to the atmosphere a likely scenario, and allowing for a first estimate regarding the efficiency of planetary mixing.} 

The question that naturally arises from these findings is how representative the inferred atmospheric composition is of the bulk of the planet, even for giant planets, where the gas dominates the mass budget. For this it needs to be understood if and how well the metals can be mixed throughout the planetary envelope. Whether this happens at all is not clear, as a gradient in metallicity (and therefore mean molecular weight) may stabilize the planetary interior against convective motions \citep[e.g.,][]{ledoux1947,chabrierbaraffee2007,lecontechabrier2012}. How a solid core or compositional gradients tend to mix throughout the planetary envelope for gas giant planets, often taking Jupiter as an example, is presently investigated \citep{vazanhelled2015,vazanhelled2016,mollgaraud2017,vazanhelled2018,muellerhelled2020,ormelvazan2020}. The results reported in these studies do not yet agree, predicting either fully mixed envelopes, or ones where a metal gradient (and core) persists in the planet. In addition to imperfect mixing also rain-out processes likely play an important role, potentially depleting both solar system and exoplanet atmospheres in metals \citep[e.g.,][]{spiegel2009,wilsonmilitzer2010}.

From the discussion above it becomes clear that the metal enrichment inferred from the atmospheric retrievals, which serve as an input for any formation analysis, is only a lower limit for the true planetary metal enrichment. As long as the metals are locked into the invisible interior of the planet homogeneously, and not selectively, this may still allow to constrain the solids' location of origin in the protoplanetary disk, as long as the atmosphere is still enriched enough for solid accretion to be the dominating factor. In the case where the relative elemental composition in the atmosphere (except for H and He) is different from the deep interior, or when the atmosphere is depleted with respect to the deeper interior to the point where it mimics the atmospheric metallicities expected from pure gas accretion, this poses a problem. Depending on how planet formation ensued, the former case may still trace the origin of the solids that were accreted towards the end of the formation process. The latter case could be resolved by checking the relative atomic abundances in the presumed metal-poor planet. If refractory atomic species are relatively abundant, this may point to a dominant accretion of solids which are mostly locked into the planet's interior. An interesting recent discussion of how to constrain planet formation in the case where the atmospheric and planetary bulk compositions differ can be found in \citet{helledwerner2021}, who come to similar conclusions. We note again that these avenues for analyzing the origin of the metals in a planet's atmosphere will be further complicated if volatile-rich gas from evaporated pebbles was accreted by the planet.

\subsection{Atmospheric evolution}
\label{subsect:atmospheric_evolution}

The atmospheric composition of an exoplanet may also evolve due to atmospheric evaporation, or by secular enrichment of infalling comets and asteroids. Evaporation is especially important for close-in low-mass planets \citep[e.g.,][]{jinmordasini2018}. The atmosphere may be partially or fully lost due to thermal or non-thermal processes, where the thermal escape separates into the regimes of Jeans escape or hydrodynamical escape, depending on the local thermal state of the atmosphere \citep[see, e.g.,][]{barman2018}. For the atmosphere to become relatively enriched in metals by evaporation two criteria have to be met. Firstly, the atmosphere needs to be of low enough mass to allow for a significant amount to be lost. Secondly, the atmospheric escape process needs to preferentially retain the heavier atmospheric species, for which the atmosphere needs to be in the Jeans escape regime \citep[e.g.,][]{bourrierdesetangs2018}. In the hydrodynamic escape regime the heavier metal species would be lost together with hydrogen. We note that this transition is gradual, however, and that there can be mass fractionation in hydrodynamic outflows as well, depending on the magnitude of the total mass flux \citep[e.g.,][]{huseager2015}. These authors also report that such outflows can selectively deplete the atmospheres of Neptune and sub-Neptune-mass planets in hydrogen over multiple Gyr, provided that the initial atmospheric mass is small enough ($<10^{-3}$ of planetary mass). For gas giant planets evaporation may thus be a less relevant process for changing the atmospheric composition. \rrp{An extreme case that is worth mentioning is the Roche lobe overflow that may affect the closest-in gas giant planets. This process could strip away the upper gaseous envelope, potentially revealing the more metal-enriched layers below. Roche lobe overflow has been discussed as the potential origin of LTT~9779~b, a planet in the hot Neptune desert \citep{jenkinsdiaz2020}. By extension, significant atmospheric evaporation may lead to similar outcomes if a compositional gradient is present in the atmosphere.}

Another possibility of atmospheric evolution is the secular contamination of the planetary atmospheres by infalling comets or asteroids. Here the frequency of cometary impacts and the persistence of the enrichment they cause in the atmosphere need to be estimated. \citet[][]{turrininelson2015} find that the additional water a comet may deposit in the visible atmosphere of the hot Jupiter HD~189733b would have to persist for 500-5000 years before being removed, assuming impacts of km-sized comets every 20-200 years, otherwise no significant enrichment is possible. A quantitative assessment of cometary enrichment requires an estimate of the persistence timescale, however. As this appear to be lacking from the literature, we present a simple first-order analysis below. We start by approximating local mixing timescales as $\tau_{\rm mix}=H_P^2/K_{zz}$, where $H_P$ is the planetary pressure scale height and $K_{zz}$ the atmospheric eddy diffusion coefficient, and find
\beq
\tau_{\rm mix} = 0.3 \left(\frac{H_P}{300 \ {\rm km}}\right)^2\left(\frac{10^8 \ {\rm cm^2/s}}{K_{zz}}\right) \ {\rm yr}.
\label{equ:tau_mix}
\eeq
A $K_{zz}$ value of $10^8 \ {\rm cm^2/s}$ within the radiative zone is well within the estimates by passive tracers reported from model calculations for HD~189733b and HD~209458b \citep{parmentier2013,agundezparmentier2014}. For self-luminous planets the correct value to be chosen is unclear, but $K_{zz}=10^5 \ {\rm cm^2/s}$ may at least be a useful lower limit \citep{ackermanmarley2001}. In the deeper regions of the atmosphere convective overshoot may lead to higher values for $K_{zz}$, smoothly transitioning towards the value expected for fully convective atmospheres as the radiative-convective boundary 
(RCB) is approached \citep[e.g.,][]{ludwigallard2002,hellingackerman2008}.

The $\tau_{\rm mix}$ estimate of less than a year in Equation \ref{equ:tau_mix} thus shows that mixing potentially proceeds on faster timescales than those quoted by \citet{turrininelson2015}, at least locally, meaning that any material added by cometary impacts should be mixed away quickly. This does not preclude a slower, homogeneous enrichment of the radiative atmosphere by cometary impacts over time. However, the question is how quickly this enrichment will be removed into the bulk interior of the planet by entrainment into overshooting convective blobs at the radiative-convective boundary. By extension, the enrichment of the visible atmosphere may be lower if the impactors deposit their metals below the radiative-convective boundary: for the bulk of Jupiter, for example, the mixing timescale has been estimated to be at most a few years \citep{debraschabrier2019}.

Assuming diffusive mixing in a 1-d atmosphere, we derive the following expression for the increase in mass fraction $X$ of a certain species in the planetary atmosphere, due to cometary impacts:
\beq
\Delta X = \frac{\dot{M}g}{4\pi R_{\rm P}^2}\frac{H _P^2}{K_{zz}}\frac{P_{\rm RCB}-P_{\rm i}}{P_{\rm RCB}P_{\rm i}}.
\label{equ:deltaX}
\eeq
Here, $\dot{M}$ is the mass accretion rate of comets, $g$ the planetary gravity, $R_{\rm P}$ the planetary radius, $H_{\rm P}$ the atmospheric pressure scale height, and $P_{\rm i}$ the average location of the destruction of impacting comets in units of pressure. $P_{\rm RCB}$ is the location of the radiative-convective boundary or, alternatively, the location where the deep mixing becomes fast enough to make the local mass fraction equilibrate to the average of the planetary interior (which is assumed to be well mixed). In any case it must hold that $P_{\rm i} < P_{\rm RCB}$ for this expression. A derivation can be found in Appendix \ref{appendix:atmospheric_evolution}.  When assuming pure-water comets, and tracking the change in the mass fraction of water, a relative enrichment of $\Delta X/X_0=3\times 10^{-4}$ to $3\times 10^{-2}$ is found, where $X_0$ is the planet's bulk water mass fraction. See Appendix \ref{appendix:atmospheric_evolution} for more details and which values to assume for the various quantities. We therefore conclude that secular cometary enrichment of the planetary atmosphere may be unlikely for giant planets, but a better modeling of the process, for example the lower boundary treatment when using a 1-d diffusion approximation, is needed.

\section{From planet composition to formation outcomes}
\label{sect:formation_inversion}

From Section \ref{sect:planet_form_complexity} it is obvious that a full inversion, from atmospheric composition to planet formation, is still a long way off. However, for qualitatively understanding the ramifications of different planet formation assumptions it would be useful to possess a tool that compares the inverted outcomes of such models. In this case the effect of a given process may be studied in isolation, allowing the user to get an intuition for the importance of a given assumption. This is in contrast to attempting to invert a full formation model, which may be either too numerically costly, or require too many parameters when compared to the limited number of observational characteristics.

In what follows, we will demonstrate such an anaylsis setup by starting with the inversion of the formation model used in \citet{oeberg2011,oebergwordsworth2019} and applying it to the compositional constraints obtained for the directly imaged planet HR~8799e. As a second step, we will introduce the effects of either including chemical evolution of the protoplanetary disk, \rrp{or the effect of pebbles that drift and evaporate in the disk}. Comparing the results of these setups for HR~8799e serves to highlight the likely importance of disk chemical evolution for its inferred migration history, \rrp{and studies whether pebble accretion may have been a likely scenario for this planet}. We end this section by suggesting other toy model setups, testing for the influence of various formation model complexities described in Section \ref{sect:planet_form_complexity}.

\begin{figure}[t!]
\centering
\includegraphics[width=0.48\textwidth]{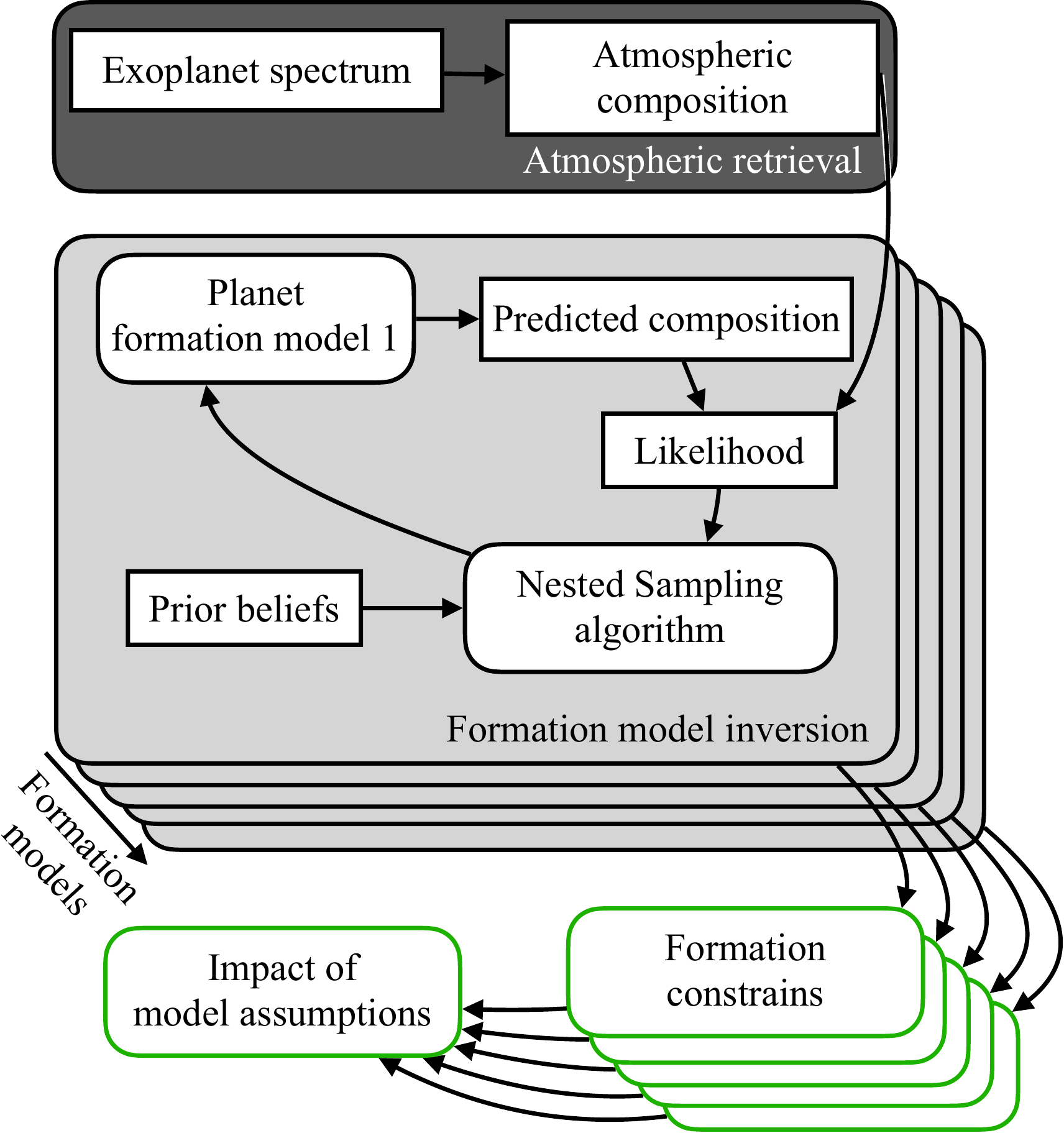}
\caption{Schematic outlining the mode of operation of a formation model inversion. Exoplanet observations first result in inferred atmospheric compositions. This composition is then used as input for the formation model inversion, which uses Nested Sampling to compare the inferred composition to the prediction of a toy planet formation model. The resulting posterior distribution of the formation parameters represent constraints of the planet formation process. 
Running inversions for various toy formation models then allows to study the impact of differing model assumptions.}
\label{fig:retrieval_framework}
\end{figure}

\subsection{Formation model inversion}
Formation models produce synthetic planets, or populations thereof, starting from a physical model and set of formation parameters. These can be, for example, the initial disk mass, the disk composition, and the starting position of a planetary embryo in a disk. When attempting to constrain planet formation based on measured planetary compositions the formation models need to be inverted, because planet composition is an outcome of the formation models. More specifically, if assuming that planetary compositions
\beq
\mathcal{C} = ({\rm O/H}, {\rm C/H}, {\rm N/H}, {\rm Fe/H}, {\rm Si/H}, {\rm P/H}, {\rm S/H}, ...)
\eeq
can be measured with a given uncertainty, we are then interested in the probability distribution of formation parameters $\vartheta$ of a formation model $\mathcal{M}$, given this measurement. Using Bayes' theorem, this can be written as
\beq
P(\vartheta|\mathcal{C},\mathcal{M}) = \frac{P(\vartheta|\mathcal{M})P(\mathcal{C}|\mathcal{\vartheta},\mathcal{M})}{P(\mathcal{C}|\mathcal{M})} .
\label{equ:form_retrieval}
\eeq
Here, $P(\vartheta|\mathcal{M})$ is the prior probability of $\vartheta$ before considering any data $\mathcal{C}$, while $P(\mathcal{C}|\mathcal{\vartheta},\mathcal{M})$ is the likelihood for observing $\mathcal{C}$, given that $\vartheta$ is true.
In practice \rrp{this may be written as, for example,}
\begin{align}
    {\rm log}P(\mathcal{C}|\mathcal{\vartheta},\mathcal{M}) & = -\sum_{i=1}^{N_{\rm species}}\left[\mathcal{C}_i - \mathcal{C}^{\mathcal{M}}_i(\vartheta)\right]^2/2\Delta \mathcal{C}_i^2 + {\rm cst} \nonumber \\
    & = -\chi^2/2 + {\rm cst},
\end{align}
where $N_{\rm species}$ is the number of measured atmospheric species, $\mathcal{C}^{\mathcal{M}}_i(\vartheta)$ is the formation model prediction of the planet abundance of species $i$, and $\Delta \mathcal{C}_i$ are the measurement uncertainties. \rrp{Here we chose a simple form of the likelihood for clarity, assuming that the measured abundances of different atmospheric constituents are independent, and follow a Gauss distribution. In general, the functional form of the log-likelihood can be arbitrarily complicated. For example, the abundance posterior of an atmospheric retrieval may be used directly, which can be approximated by, say, a Gaussian mixture model. In our application in Section \ref{sect:formation_inversion} we chose an intermediate step, accounting for the covariance between the atmospheric oxygen and carbon content.} We note that such an inversion process does not necessarily only need to consider elemental abundances as input measurements. Any observed property of a planet, such as its orbital parameters, could in principle be included in this analysis, as long as it is predicted by a formation model.

In practice, we will compute samples of the target distribution $P(\vartheta|\mathcal{C},\mathcal{M})$ by numerically integrating the numerator of the right-hand side of Equation~\ref{equ:form_retrieval} using the so-called Nested Sampling method \citep{skilling2004}. In short, Nested Sampling is a Monte Carlo technique to integrate functions in highly-dimensional model parameter spaces. Here it is the space spanned by the formation parameters. When integrating the numerator of 
Equation~\ref{equ:form_retrieval}, the integral value resulting in the so-called model evidence, Nested Sampling will automatically generate samples of our target distribution. In principle model evidence ratios allow to distinguish between different formation models. However, as long as we cannot invert full state-of-the-art formation models, this may be possible only when considering sets of assumptions that lead to wildly different outcomes, and where one clearly represents a better fit. We use the \texttt{PyMultiNest} \citep{buchnergeorgakakis2014} package for inverting formation models, which is a Python wrapper of the \texttt{MultiNest} code \citep{ferrozhobson2013}.
A schematic illustrating our approach is shown in Figure 
\ref{fig:retrieval_framework}.

\subsubsection{Inverting the \"Oberg et al. (2011) model}
\label{subsect:inverting_oeberg}
As a first toy model we will use arguably one of the simplest formation models conceivable when aiming at studying the usefulness of atmospheric C/O constraints. For this we follow the setup presented in \citet{obergmurray-clay2011} and \citet{oebergwordsworth2019}, assuming their static protoplanetary disk model as described for the young solar nebula, \rrp{but with the icelines and composition adapted to the planet system of interest}. For every volatile species we define a constant mass fraction in relation to the total disk mass. Inside of its iceline the volatile species is in the gas phase, outside it is in the solid phase. This leads to the well-known step-like behavior of the C/O values in the solid and gaseous phase of the disk, as shown in Figure \ref{fig:oeberg_steps}. More details on our implementation can be found in Appendix \ref{appendix:disk_model_Oeberg}.

We then assume that the planet formation process can be fully described by a set of four parameters, which ultimately allow to map to the bulk composition of the planet:
\beq
\label{equ:formation_parameters}
\vartheta = (M_{\rm P}, M_{\rm solid}, a_{\rm solid}, a_{\rm gas}),
\eeq
where $M_{\rm P}$ is the total planet mass, and $M_{\rm solid}$ is the mass of the solids (refractory species and volatile ices) accreted by the planet.  It holds that $M_{\rm P} = M_{\rm gas} + M_{\rm solid}$, which we use to determine the amount of gas a planet accreted. The parameters $a_{\rm solid}$ and $a_{\rm gas}$ denote the orbital distances where the solids and gas were accreted, respectively.

For a given value of $\vartheta$ our setup then uses the disk model to determine the planet's accretion location with respect to the disk icelines. Species in the gas phase will be used to determine the composition of the accreted gas; the analogous is done for the solids. Together with the mass fractionation between solids and gas in the planet this determines the planet's composition $\mathcal{C}$.

\subsubsection{Adding chemical time evolution}
\label{subsect:chemical_time_evo_inversion}

The \citet{oeberg2011} disk setup is convenient for conceptually studying the usefulness of C/O. As discussed in Section \ref{sect:planet_form_complexity} there are many complicating factors that make a true inversion from C/O to planet formation parameters extremely challenging. In our second toy model setup we single out one of these processes, and study how chemical evolution in the protoplanetary disk changes inferences when compared to the static disk model.

\begin{figure}[t!]
\centering
\includegraphics[width=0.4975\textwidth]{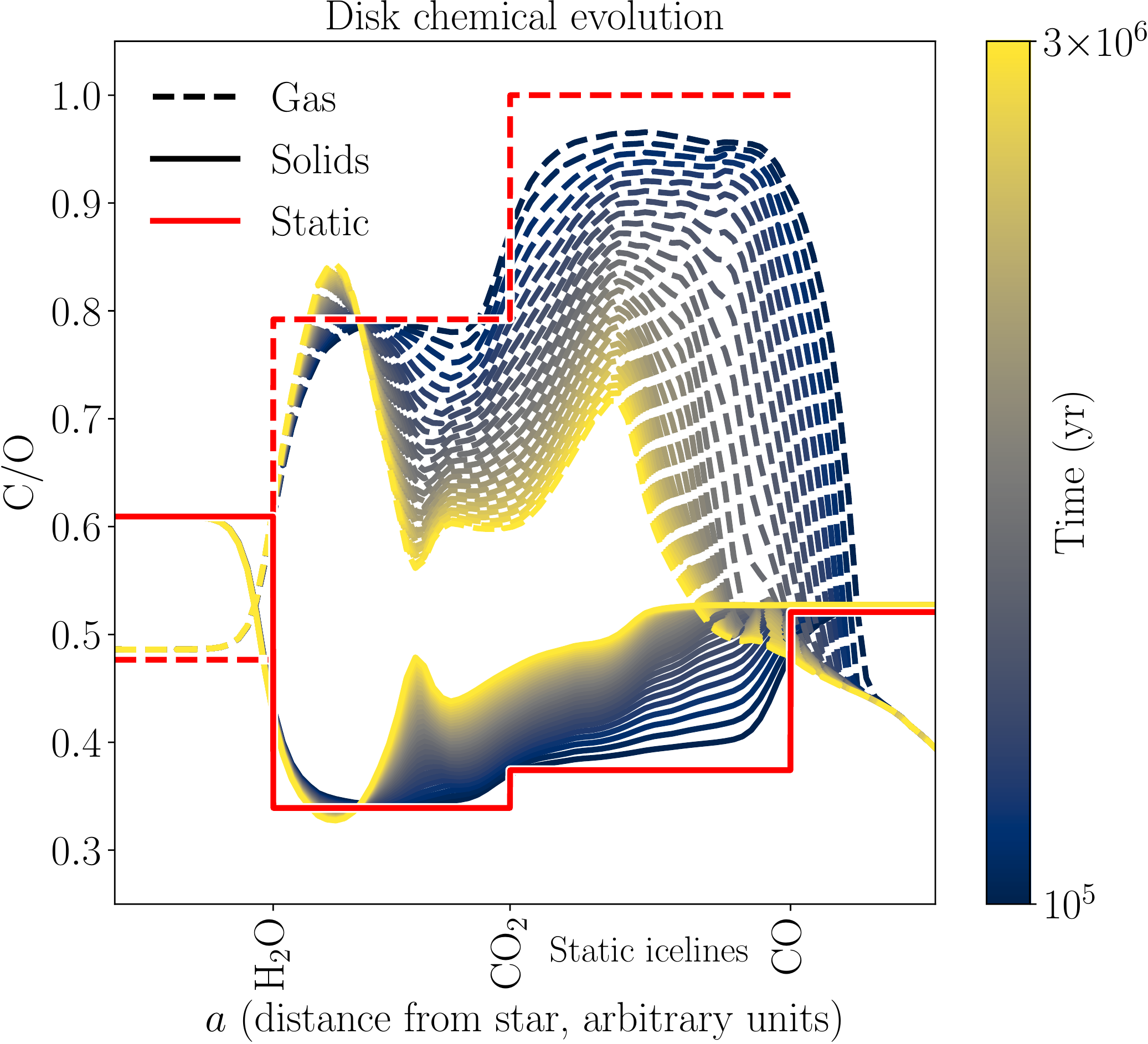}
\caption{C/O values in the protoplanetary disk's gas and solid phases, shown as dashed and solid lines, respectively. C/O ratios shown in red are obtained from the \citet{oeberg2011} disk model (static disk composition). C/O ratios shown in colors changing from yellow to dark blue are obtained from the model including the chemical evolution of the protoplanetary disk.}
\label{fig:oeberg_vs_time_evo}
\end{figure}

As discussed before, chemical reactions may process the initial disk abundances over time, shifting carbon and oxygen atoms to different chemical species. For example, CO, which is very volatile, can be removed from the gas by surface reactions, processing it into less volatile CO$_2$ \citep{molyarovaakimkin2017,2018A&A...613A..14E,2018A&A...618A.182B,schwarzbergin2018}. If a large fraction of CO gas in a given region of the disk mid-plane undergoes such processing into CO$_2$ ice, then an amount of elemental C and O equivalent to that initially carried in CO will be processed from the gas into the ice in this region. This is one example of how chemical processing in disks can alter the elemental partitioning of, for example, C and O, in the gas and ice. These processes are thus expected to influence inferences on the location of planet formation as first discussed in \citet{2018A&A...613A..14E}.

To study the effect of disk chemical evolution in formation inversions, we replaced the static disk abundance model with a time-dependent model. For this we calculated the evolution of the disk chemical composition using the ANDES code, which describes a quasi-stationary 2-d axis-symmetric protoplanetary disk \citep{2013ApJ...766....8A,molyarovaakimkin2017}. ANDES solves for the time-dependent chemical composition of the disk with a detailed description of grain surface and gas-phase reactions, see Appendix \ref{appendix:disk_model_ANDES} for more details. For the initial disk setup we use the abundances given in Table~\ref{tab:disk_setup}, that is, equal to the disk abundances used for our static disk model. Although the chemical model includes other elements, we only consider H, He, C, O, and N-bearing species in the calculations. The refractories are considered to be chemically inert, and condensed at all times. An example for the resulting time evolution of the C/O ratio in the disk is shown as the yellow to dark blue lines in Figure \ref{fig:oeberg_vs_time_evo}. The process of planet formation is then modeled in the same way as in the static disk case, but with the difference that the composition of the accreted gas and solids are taken from the disk's chemical evolution calculations. Formally this also allows to add the formation time as an additional parameter to be constrained, but here we chose to initially only study and compare the inversion outcomes for different times during chemical evolution.

\subsubsection{\rrp{Studying the effect of pebble accretion}}
\label{subsect:pebble_inversion}

\rrp{In addition to the two setups described above, we study the effect of pebbles drifting and evaporating in the protoplanetary disk, and their accretion, as a third scenario. In general, pebbles will quickly drift towards the central star in an unperturbed disk, because of the torque exerted by the head wind of the disk gas. This wind is caused by the radial pressure support of the gas, which makes it orbit the star at  sub-Keplerian speeds. Therefore, unless there are local pressure enhancements that trap pebbles (caused by, e.g., a planet), pebbles will drain into the inner parts of the disk quickly, releasing copious amounts of volatiles into the gas phase when crossing their respective icelines \citep[e.g.,][]{boothclarke2017,boothilee2019,schneiderbitsch2021}. Therefore, in addition to accreting pebbles directly, this process can be crucial for setting the composition of forming planets by gas accretion.

For modeling the effect of pebbles on the formation inversion of HR~8799e, we fed disk compositional structures from the {\tt chemcomp} model \citep{schneiderbitsch2021} into our inversion framework. In short, {\tt chemcomp} solves for the pebble growth, drift, and evaporation in a viscously evolving disk. In addition, it includes a full planet formation model in the pebble accretion paradigm, handling planet-disk interactions such as gap opening and planet migration. Because {\tt chemcomp} is too slow for inverting it directly, we use its pebble drift and evaporation prescription in an unperturbed disk to obtain a disk compositional structure as a function of time, to which we then apply our inversion framework, indentical to the treatment of the disk's chemical evolution.}

\subsection{Application of toy model inversions to HR~8799e}
\label{subsect:formation_inversion_hr8799e}

As discussed in Section \ref{sect:introduction}, deriving accurate and precise C/O ratios from current exoplanet observations is challenging, but this will likely change with {\it JWST}. Using high-resolution spectrographs, or the interferometric {\it GRAVITY} instrument at the VLT, the community is already starting to derive precise C/O values using ground-based instruments. This has to do both with the data, but also the use of state-of-the-art retrieval techniques \citep{brogiline2018,gandhimadhusudhan2019b,mollierestolker2020,pelletierbenneke2021,linebrogi2021}. Below we will make use of the atmospheric composition derived for HR~8799e from {\it GRAVITY} observations \citep{mollierestolker2020}, and try to constrain how its derived formation history changes when disk chemical time evolution \rrp{or pebbles} are included.

\rrp{In order to run the formation inversion for HR~8799e, a disk elemental composition needs to be assumed for HR~8799. HR~8799 is a $\lambda$-Bo\"otis-type star; this means that the abundances measured for its iron-peak elements are subsolar, with values of $\rm [Fe/H] = -0.55\pm 0.10$ \citep{sadakane2006} or $\rm [Fe/H] = -0.52\pm 0.08$ \citep{wangwang2020} having been inferred for iron specifically. A similar depletion is expected for Mg, Si, and other massive iron-peak elements, while the abundances of elements typically found in volatile species (C, N, O) are expected to be close to solar \citep[e.g.,][]{paunzen2004}. Indeed, the latest analysis of \citet{wangwang2020} inferred [C/H]\ $ = 0.11 \pm 0.12$, [O/H]\ $=0.12\pm 0.14$, and (C/O)/(C/O)$_\odot = 0.96 \pm 0.19$, which are all consistent with solar, but slightly enriched in C and O. According to \citet[][]{wangwang2020}, the most likely explanation for the observed composition of HR~8799 is recent accretion of volatile-rich material onto the outer layers of the star, for example from an evaporating hot Jupiter, or of volatile-rich ices scattered into the inner system by the four HR~8799 planets. We will therefore use the composition of the Solar System from \citet{oebergwordsworth2019} as our nominal abundance model, because it could be unlikely that the star has a bulk elemental composition identical to its observed photospheric values. When relevant, we will also report on how our results change if taking the $\lambda$~Boo abundances of HR~8799 at face value, however. The iceline positions in the HR~8799 disk are set to the values derived from the ANDES chemical evolution model, at $t = 0$.}

\begin{figure*}[t!]
\centering
\includegraphics[width=1.00\textwidth]{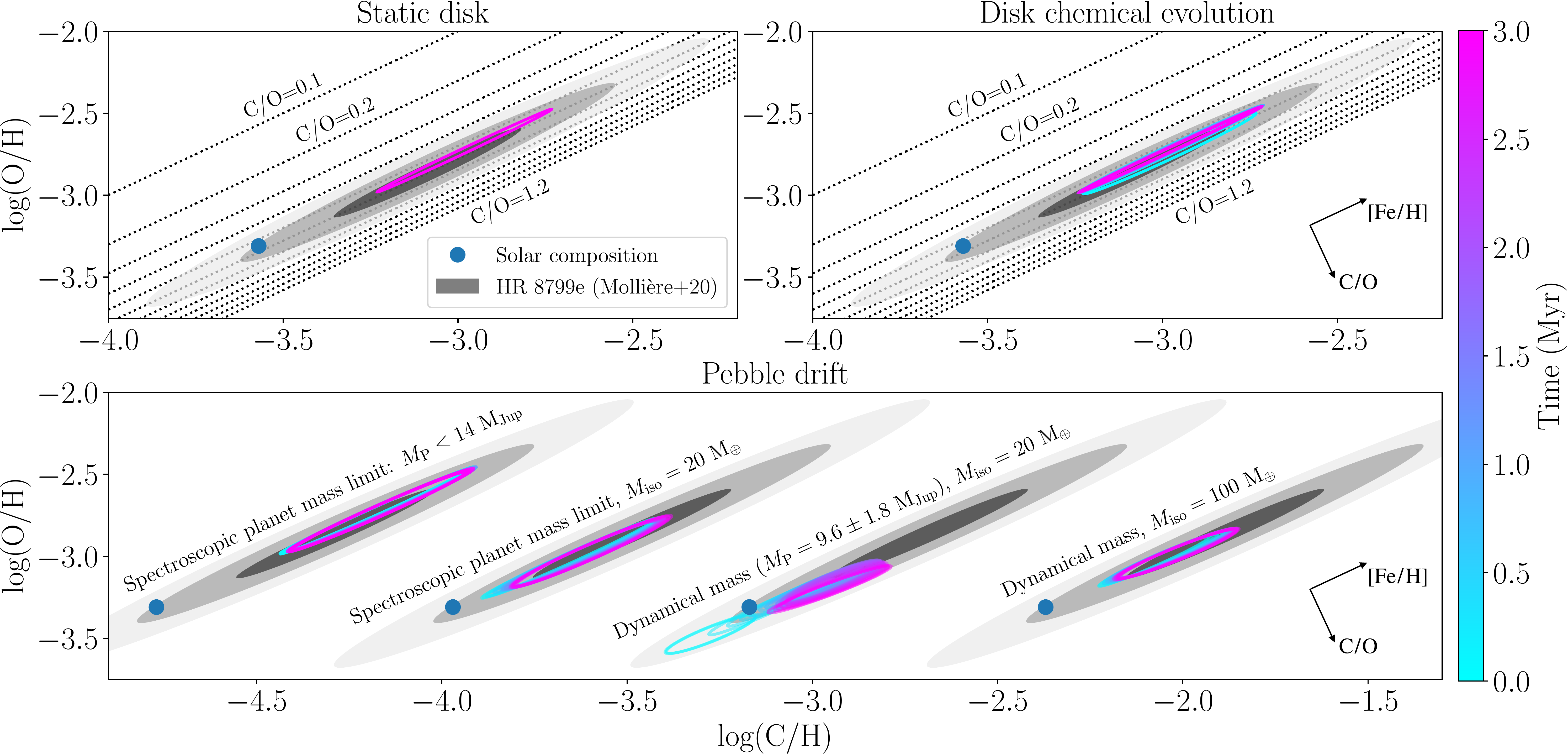}
\caption{C/H and O/H distribution of HR~8799e (black ellipses) as obtained from the GRAVITY retrievals reported in \citet{mollierestolker2020}. The three ellipses denote the 1,2,3-sigma uncertainties of the C/H, O/H distributions. \rrp{Colored ellipses show the distribution of values (1-sigma) resulting from running formation inversions as a function of time for different disk scenarios.} The slanted dashed lines denote (C/H, O/H) value pairs of constant C/O from 0.1 to 1.2, in steps of 0.1. The solar C/H, O/H value is shown as a blue filled circle. \rrp{Arrows indicate the direction of increasing atmospheric metallicity and C/O.} \rrp{{\it Upper left panel:} formation inversion using a static disk model. {\it Upper right panel:} formation inversion taking into account the chemical evolution of the disk. {\it Lower panel, from left to right:} formation inversions in cases including pebble drift and evaporation. Case (i): replacing the static disk model by the pebble disk model. Case (ii): same as Case (i), but additionally placing a prior on the accreted solid mass, corresponding to a 20-$\rm M_\oplus$ upper limit defined by the pebble isolation mass. Case (iii): same as Case (ii), but including a tighter prior on the mass of HR~8799e, based on a dynamical mass estimate. Case (iv): same as Case (iii), but using a larger value for the pebble isolation mass (100~$\rm M_\oplus$). An arbitrary offset has been applied to the ellipses of Cases (i) to (iv), for clarity.}}
\label{fig:CoHOoHs_GRAVITY}
\end{figure*}

For HR~8799e we make use of the atmospheric retrieval results \rrp{reported in \citet[][]{mollierestolker2020}}. Of relevance for the formation inversion are the derived values for the planetary mass, as well as the \rrptwo{atmospheric} metallicity and C/O ratio. \rrptwo{As the mass is spectroscopically determined, it has large error bars. However, it still results in a constraint on the total amount of solids that the planet incorporated, which can also be estimated from multiplying the atmospheric metallicity by the planetary mass. More specifically, multiplying the planet mass with the inferred atmospheric metallicity results in a lower limit on the metal mass (in solid or gaseous state) that a planet accreted, which may be dominated by solids in cases of high metallicities.} We used the actual posterior distributions on planetary gravity and radius from the spectral retrievals to construct the inversion prior for the planetary mass (effectively corresponding to a 1-$\sigma$ upper limit of 14~${\rm M_{Jup}}$). \rrp{We also study the effect of using a tighter mass constraint in the pebble inversion scenario.}

Converting the atmospheric C/O ratio for use in the inversion method requires special care. In the spectral retrievals the metallicity is used as a free parameter to scale all elemental abundances except H and He, after which the C/O ratio is set by scaling O. In the formation model the formation location as well as relative gas-to-solid accretion fraction sets the O, C, and refractory metal content, from which a C/O ratio can be calculated. Therefore the C content is no longer strictly coupled to the refractory metal content, in contrast to the spectral retrievals which we use as input for our formation retrievals. This inconsistency has to be kept in mind when we use the C/O and metallicity of the spectral retrieval to obtain atmospheric C/H and O/H values, and compare these to the C/H and O/H predicted by the formation model. In general, independently constraining C/H and O/H in atmospheric retrievals is the better avenue for running formation inversion studies. \rrp{We note that it is also important to account for the amount of oxygen that has been sequestered into atmospheric clouds. Because the C/O constraints from \citet[][]{mollierestolker2020} include this effect, the atmospheric retrieval results can be used without modification. For C/O constraints from retrievals that constrain absorber abundances independently, corrections would need to be applied.}

\begin{figure*}[t!]
\centering
\includegraphics[width=0.458\textwidth]{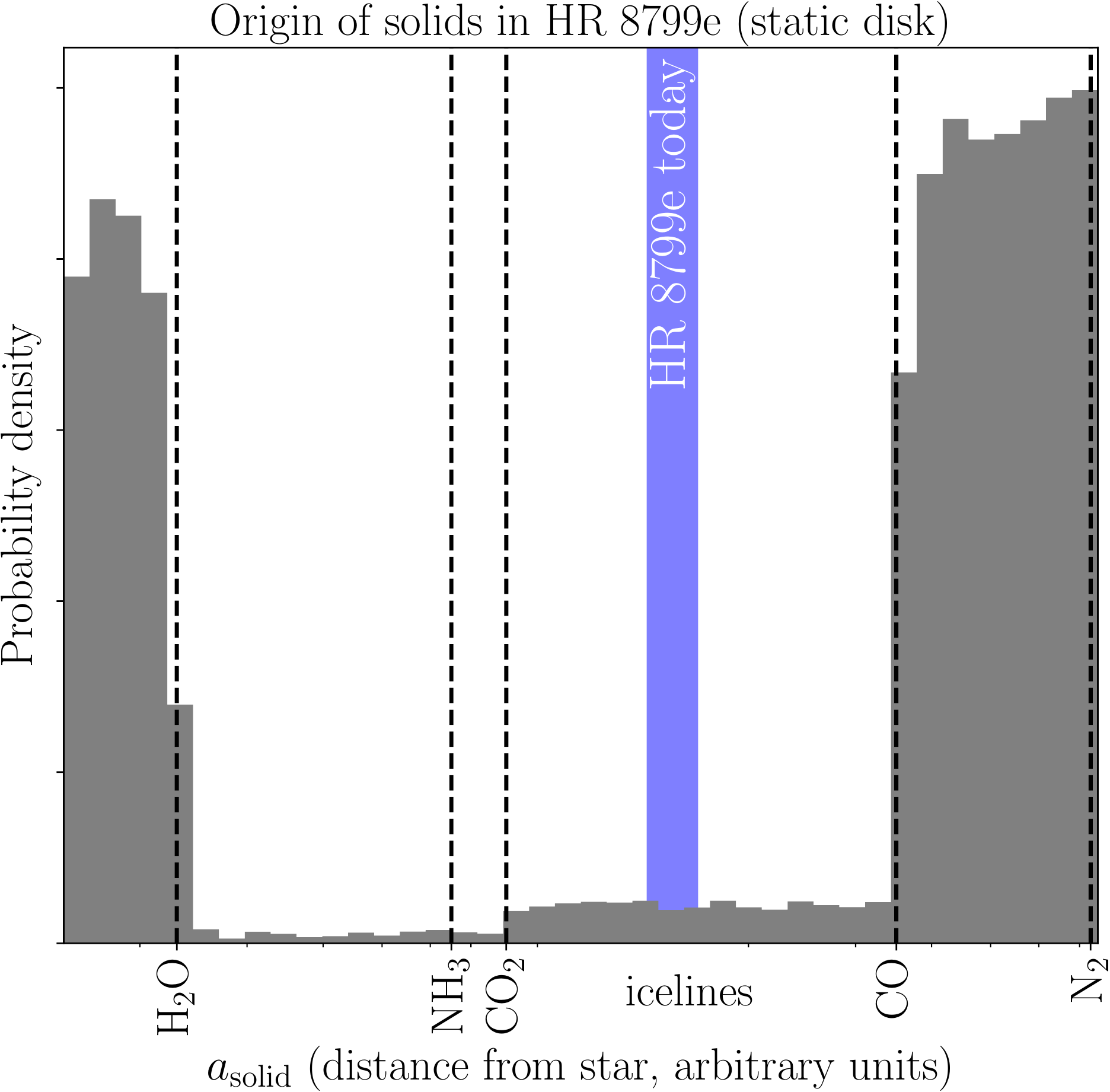}
\includegraphics[width=0.503\textwidth]{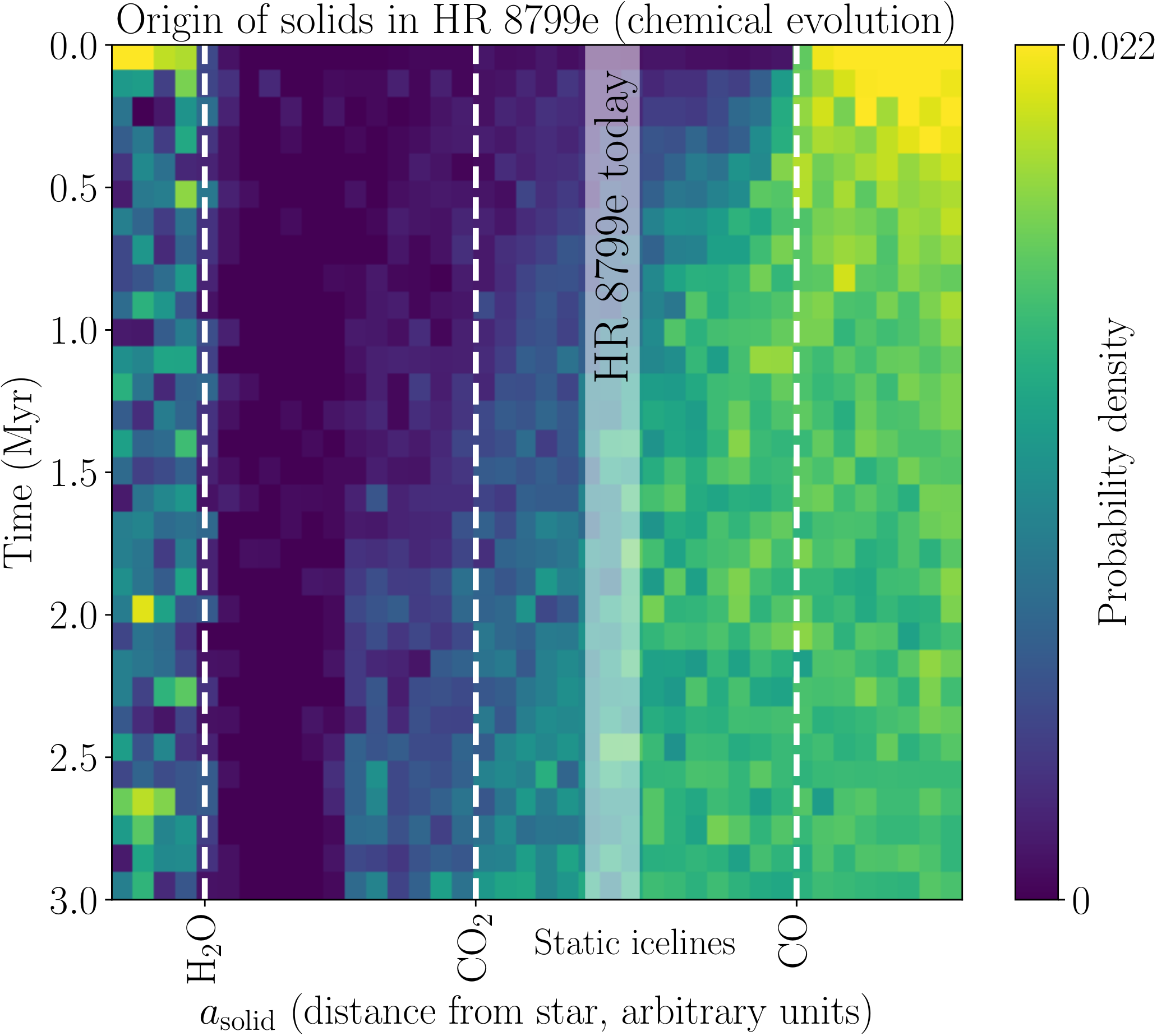}
\caption{{\it Left panel:} posterior distribution of the location where the solids in HR~8799e were accreted or originate when using the static disc composition in the formation model. Because the planet has an increased metallicity and stellar C/O ratio its enrichment is likely dominated by solids which originate from outside the CO iceline, as the ices then incorporate all major C- and O-bearing species. HR~8799e's current orbital distance is shown as a blue vertical line, where the iceline positions were computed from chemical disk models around a young HR~8799 host star, see description in text. {\it Right panel:} formation inversion of HR 8799e including the chemical evolution of the protoplanetary disk. The iceline positions of \ce{NH3} and \ce{N2} have been omitted for clarity. The x-axis has been log-scaled in both panels.}
\label{fig:solid_orig_HR8799e}
\end{figure*}

Moreover, sampling the C/O and metallicity posterior of the spectral retrieval leads to tightly correlated C/H and O/H values, and we take this into account by fitting it with a two-dimensional Gaussian distribution. The distribution's covariance matrix is then used to describe the uncertainties of C/H and O/H during the inversion process. If this was not done the independent uncertainties in C/H and O/H (as obtained from the diagonals of the covariance matrix) would allow for a spread in C/O values much larger than obtained from the spectral retrieval, rendering a formation inversion meaningless. Taking the spectral retrieval results from \citet{mollierestolker2020} we used ${\rm [Fe/H]}=0.48\pm 0.27$ and ${\rm C/O}=0.60\pm 0.075$. The resulting 2-d distribution of C/H and O/H is shown in Figure~\ref{fig:CoHOoHs_GRAVITY}.

Lastly, because our adapted \citet{oebergwordsworth2019} setup for the disk abundance led to a slightly sub-solar C/O ratio (0.52 instead of 0.55), we scaled the planetary value of 0.6 by 0.52/0.55 prior to generating the planetary C/H, O/H distribution as input for the formation inversion process. This is done to conserve the relative distance in C/O, with respect to solar abundances, of the input planetary composition.

\subsubsection*{Static disk chemistry}
We start the formation inversion for HR~8799e using the disk model \`a la \citet{oeberg2011,oebergwordsworth2019}. The \rrp{magenta} ellipse in \rrp{the upper left panel of} Figure \ref{fig:CoHOoHs_GRAVITY} shows the distribution of sampled C/H, O/H pairs of the formation inversion, indicating a good fit. The most interesting result from the formation inversion is shown in the left panel of Figure \ref{fig:solid_orig_HR8799e}: here we plot the 1-d posterior of $a_{\rm solid}$, that is the location where the solids that are enriching the planet originated or were accreted.

The inversion process finds a clear preference for the solids to stem from outside the CO iceline, or from within the \ce{H2O} iceline. This is intuitively easy to understand: because the spectral retrieval resulted in a superstellar atmospheric metallicity, and a C/O ratio consistent with stellar, this means within the \citet{oeberg2011} model, that the planet must have accreted solids of stellar C/O ratio. \rrp{The metal content of the gas, being stellar or sub-stellar, is not high enough to offset the C/O ratio set by the accretion of solids.} Accreting solids of stellar C/O is possible at orbital distances outside of the icelines of all major carbon and oxygen carrying species, the outermost being CO (see Figure \ref{fig:oeberg_steps}). For HR~8799e's current orbital position $\sim$15~au \citep{wanggraham2018}, this could mean that the planet underwent some orbital migration after solid accretion, as the CO iceline for a young A5 host star such as HR~8799 is expected to be around 35~au. To illustrate this, we overplot today's orbital location of HR~8799e in the left panel of Figure \ref{fig:solid_orig_HR8799e}.

Alternatively, due to a high organic carbon content of the refractories in our toy model, a roughly stellar C/O ratio is also attainable if the planet formed within the water iceline, and then migrated or scattered outward to its current orbital position. This formation channel for distant giant planets was proposed, for example, in \citet{marleaucoleman2019}. While this is an intriguing result, we stress that it is dependent on the disk compositional model we assume, and the formation model used in general. We also note that the high carbon content of the refractories in the inner disk may be unlikely, see our discussion in Section \ref{sect:planet_form_complexity}. The 1- and 2-d projection of the full posterior of the formation inversion is shown in the left panel of Figure~\ref{fig:hr8799_form_full_posterior}, and discussed in Appendix \ref{app:hr8799e_full_posterior}.

\rrp{We also carried out inversions using the $\lambda$ Boo composition of the star for the disk. This was done by increasing the oxygen and carbon abundance by 30~\%, and decreasing the iron and silicate content of the refractory material by 70~\%. The oxygen no longer bound in silicates was added to \ce{H2O}, which is the dominant reservoir of oxygen in the protoplanetary disk. \ce{CO}, the second most abundant oxygen reservoir should not change, because the carbon content of the disk is not changed when applying the depletion of the iron-peak elements. Finally, the oxygen abundance is increased until C/O=0.54 is reached, which is the value reported in \citet{wangwang2020}. We find that the most likely location of origin of the accreted solids is still outside the CO iceline. The formerly second likely location, inside the \ce{H2O} iceline, vanished: the solid C/O there is exclusively set by the refractory species, which have a much higher C/O value of 2.6 now, because of the strong silicate depletion. The associated posterior is shown in the right panel of Figure~\ref{fig:hr8799_form_full_posterior}, in Appendix \ref{app:hr8799e_full_posterior}.}

\subsubsection*{Chemical disk evolution}
Next we analyze how robust the above findings are when adding chemical evolution of the disk composition. We thus ran a formation inversion of HR~8799e using the formation model that included chemical evolution. In practice this was done by inferring the planet formation parameters using the composition of the ANDES disk chemical models, as a function of time. ANDES computes the abundances as a function of altitude above the midplane. We used the resulting surface densities of the disk to determine the composition of the gas and solids. We assumed a young (1~Myr) host star at HR~8799's current mass and $L=3.58$~$\rm L_\odot$ \citep{yorkebodenheimer2008}, with a disk that produces an accretion luminosity of $0.233$~$\rm L_\odot$ \rrp{(corresponding to $10^{-8} \ {\rm M_\odot \ yr^{-1}}$)}, where $\rm L_\odot$ and $\rm M_\odot$ are the solar luminosity and mass. For a given inversion at time $t$ we assumed that the disk composition is fixed at the value that the chemical evolution predicts at that time. This is an approximation, because it implicitly assumes that planet formation happens over a characteristic timescale $<10^5$ yr, which is chosen as the time step between the snapshots of the disk composition. Because the goal of the present exercise is to study the zeroth order effect that chemical evolution may have, we deem this approximation acceptable. Future applications could incorporate the time of formation as another free parameter, also assuming (or trying to infer) the duration of the planet formation process.

The results of the inversion including disk chemical evolution are shown in the \rrp{upper right panel of Figure~\ref{fig:CoHOoHs_GRAVITY}, indicating a good fit of the atmospheric composition}. The right panel of Figure \ref{fig:solid_orig_HR8799e} shows the posteriors on the location where the planet accreted its solids (or, alternatively, where these solids originated in the disk) \rrp{as a function of time}. For reference, the iceline positions of the static disk model are indicated as well.

\begin{figure*}[t!]
\centering
\includegraphics[width=0.49\textwidth]{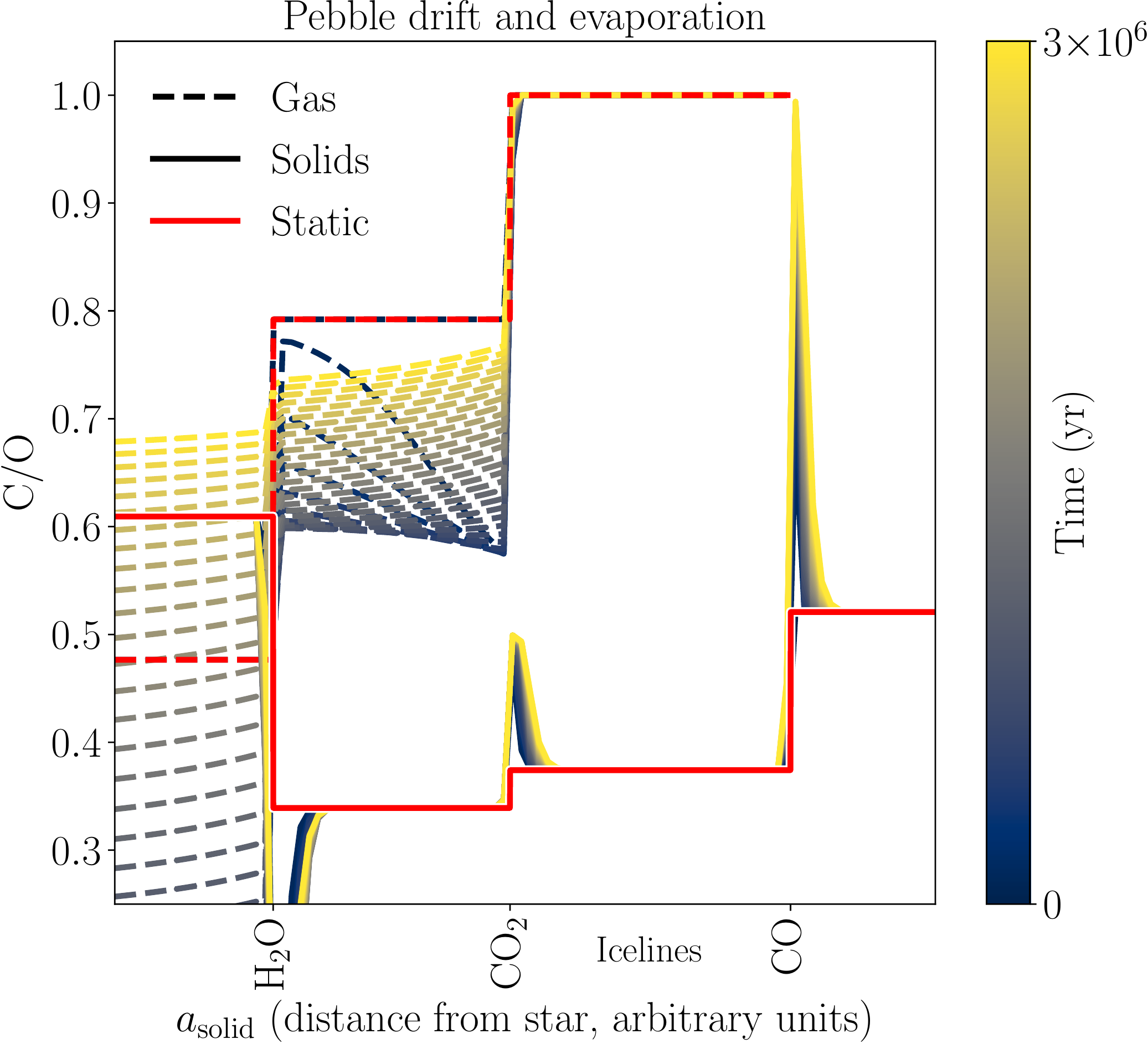}
\includegraphics[width=0.495\textwidth]{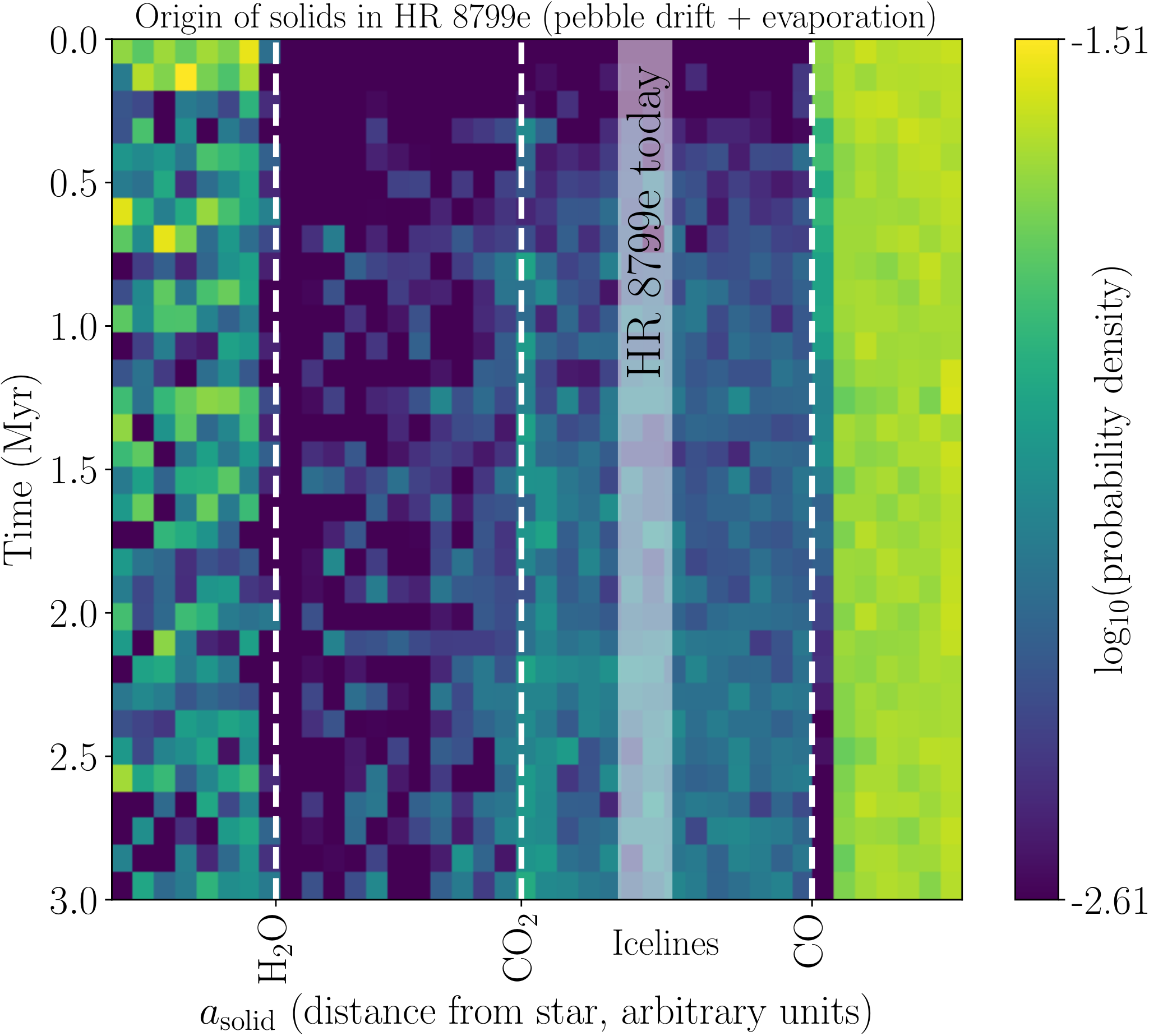}
\caption{\rrp{{\it Left panel:} C/O in the disk's gas and solid phase when incorporating the effect of pebbles drifting to the inner disk, and their evaporation at the icelines. {\it Right panel:} time-dependent 1-d posterior of $a_{\rm solid}$ for the formation inversion of HR~8799e in the pebble drift scenario, Case (i). The iceline positions of \ce{NH3} and \ce{N2} have been omitted for clarity, and the x-axis has been log-scaled.}}
\label{fig:solid_orig_HR8799e_pebble}
\end{figure*}

To understand the results of the inversions with chemical evolution it is useful to reconsider the underlying C/O distribution in the disk as a function of time, shown in Figure \ref{fig:oeberg_vs_time_evo}. For $t>0$ it is seen that the disk gas C/O value outside the \ce{CO2} iceline decays, while the C/O of the solid component increases. This is because CO is converted into CO$_2$ ice on the surfaces of dust grains outside the iceline of CO$_2$ over time. The conversion rate depends on the CO abundance in the ice, which drops rapidly inside the CO iceline. So while the reaction rate increases with temperature, the conversion is more efficient right inside the CO iceline \citep{2018A&A...618A.182B}. Thus, the process occurs first for larger disk radii, and later for smaller radii, and drives the solid C/O towards the stellar value also inside the static CO iceline. We note the the ANDES model also included the formation of \ce{CH3OH} ice.

Because HR~8799e is found to have a C/O ratio similar to the stellar one, this means that the region for its most likely formation (or the region of origin of its accreted solids) expands inwards over time to include smaller disk radii. This effect is clearly visible in the right panel of Figure \ref{fig:solid_orig_HR8799e}. We therefore confirm the findings presented in \citet{mollierestolker2020}, where it was argued that processing CO gas into CO$_2$ ice may have a significant effect on the formation location of HR~8799e. As stated in \citet{mollierestolker2020}, this also has consequences for how strongly HR~8799e may have migrated to reach its present-day orbit. If chemical evolution was significant in HR~8799e's natal protoplanetary disk, and if the exoplanet formed late enough, it may have migrated much less (or not at all) than in cases where it formed early.

In general, our findings emphasize the importance of disk chemical evolution for planet formation that has been reported in \citet{2018A&A...613A..14E}. It also shows that any analyses that try to infer planet formation based on atmospheric compositions should compare the relevant chemical timescales to the timescales of planet formation.

\rrp{Similar to the static disk case, assuming $\lambda$~Boo-type elemental abundances for the chemical evolution is not expected to change the results significantly. Inreasing the carbon and oxygen abundance by 30~\% is within the modeling uncertainties of the disk chemistry, and the additional oxygen going into \ce{H2O} due to the silicate depletion is irrelevant to the evolution of the CO iceline. In the inner part of the disk, within the \ce{CO2} iceline, water ice is slowly destroyed to form \ce{CO2} ice, which raises the solid C/O in the inner disk over time, similar to the CO condensation within the static \ce{CO} iceline. The amount of available \ce{CO2} that can be formed is independent of the \ce{H2O} ice fraction to first order, and the timescale over which this happens is set by the cosmic ray ionization rate, so it is independent of the water concentration.}

\subsubsection*{\rrp{Pebble drift and evaporation}}

\rrp{In this section we model the effect of pebble drift and evaporation on the formation inversion of HR~8799e. This process can be crucial for setting the composition of forming planets. When neglecting pebble drift, planets whose atmospheric metal content is set by gas accretion are generally expected to have sub-stellar metallicities and super-stellar C/O values. In contrast, planets with an atmospheric metal enrichment dominated by solid accretion may have have super-stellar metallicities, but sub-stellar C/O ratios \citep[e.g.,][]{oeberg2011,madhusudhan2014,mordasinivanboekel2016,madhusudhanbitsch2017}. In the case of pebble drift, however, evaporation of pebbles inside of the CO, \ce{CO2}, and potentially the \ce{CH4} icelines can lead to disk gas that is significantly enriched in these species, allowing for super-stellar metallicities and C/O ratios in the disk's gas phase, and therefore in the atmospheres of planets \citep[e.g.,][]{boothclarke2017,schneiderbitsch2021,schneiderbitsch2021b}.

\begin{figure}[t!]
\includegraphics[width=0.47\textwidth]{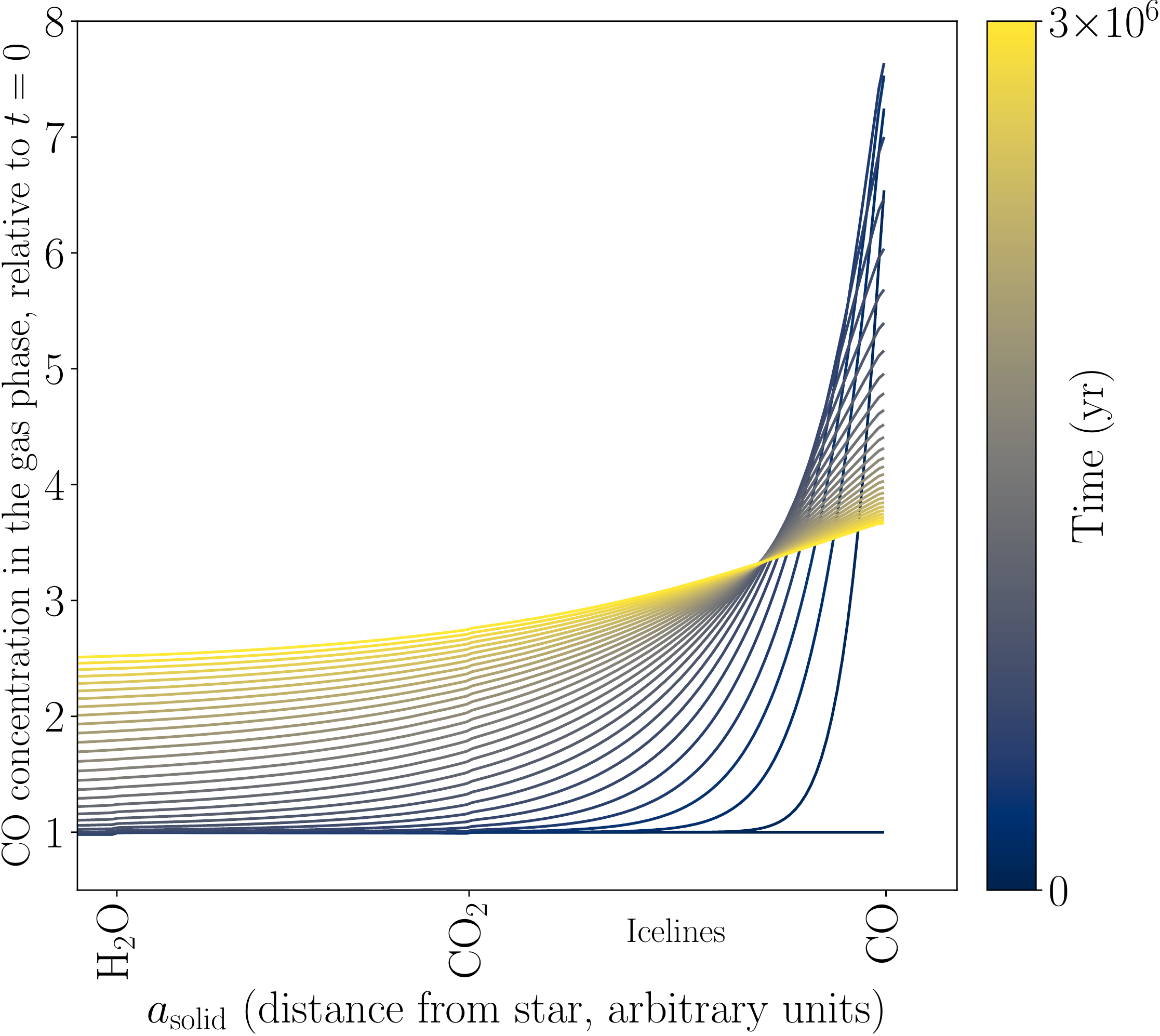}
\caption{\rrp{Evolution of the CO concentration of the disk gas, normalized by the initial CO concentration, in the pebble drift and evaporation scenario. The earliest times are characterized by a spike in CO close to the CO iceline, due to pebble evaporation, followed by its viscous spreading.}}
\label{fig:CO_increase}
\end{figure}

For setting up the {\tt chemcomp} pebble disk model, we used the same initial disk surface density and temperature structure as for the disk's chemical evolution case. Likewise, the initial disk composition was fixed to the one described in Table~\ref{tab:disk_setup}. For the disk viscosity we chose an intermediate value of $\alpha = 5\times 10^{-4}$, where $\alpha$ is the usual dimensionless diffusion coefficient, in units of $c_{\rm s}H$, where $c_{\rm s}$ is the local mid-plane sound speed and $H$ the disk's pressure scale height \citep{shakurasunyaev1973}. This value is consistent with observational data on turbulence in protoplanetary disks, suggesting $\alpha$ of the order of $10^{-3}-10^{-4}$ \citep{2016ApJ...816...25P,2017ApJ...843..150F,2018ApJ...856..117F}. The disk viscosity is a key parameter for the pebble problem, with smaller values of $\alpha$ leading to larger pebbles, thus generally faster inward drift, and longer persistence time scales of the gas locally enriched by pebble evaporation.
All solid material is considered to be in the form of pebbles, with an initial particle size of 1~$\mu$m, which then evolve by growth and drift \citep[e.g.,][]{birnstielklahr2012}.

The disk's resulting C/O values in the solid and gas phase are shown in the left panel of Figure~\ref{fig:solid_orig_HR8799e_pebble}. At $t=0$ the disk C/O values reproduce our static setup. At larger times, however, the effect of drifting pebbles becomes noticeable very quickly. Pebbles drifting across the CO iceline will start enriching the gas phase in CO (also see Figure \ref{fig:CO_increase}). Some of this gas diffuses outward again, condensing on the inward drifting pebbles, and increasing the pebble C/O value to unity just outside the CO iceline. The same effect is visible just outside the the \ce{CO2} and \ce{H2O} icelines, where the solid C/O values reach 0.5 and 0, respectively. Away from the icelines the C/O  of the solids remains largely unchanged, however. At the same time we note that the solid surface density will drop significantly over the simulated time due to pebble drift, by up to two orders of magnitude, while the gas surface density only drops by less than one order of magnitude. Inside the \ce{CO2} iceline the C/O ratio of the gas immediately drops at $t>0$ as \ce{CO2} evaporates off the inward drifting pebbles. At later times the gas' C/O starts rising again as the CO gas that has evaporated off the pebbles inside the CO icelines reaches the inner disk regions, due to the disk's viscous evolution. An analogous evolution can be observed for the disk gas inside the \ce{H2O} iceline; the gas' C/O value first drops significantly due to the water evaporating off the pebbles, but rises again at later times as gas enriched in \ce{CO2} and CO viscously spreads inwards.

For studying the effect of pebble accretion, drift, and evaporation, we investigated four scenarios with our formation inversion setup. Case~(i) is simply applying the disk compositional model, as determined by the pebble drift and evaporation framework, in the formation forward model. Case~(ii): like (i), but putting an upper limit of 20~$\rm M_{\oplus}$ on the mass that can be accreted as solids, accounting for the concept of the pebble isolation mass (see Section~\ref{subsect:planet_formation}, and \citealt{bitschmorbidelli2018}). Case~(iii): like (ii), but replacing the upper mass limit on the planetary mass from the spectroscopic retrieval \citep[$M_{\rm P}<14 \ {\rm M_{Jup}}$,][]{mollierestolker2020} with a tighter prior from the dynamical mass estimate reported in \citet{brandtbrandt2020}, that is $M_{P} = 9.6_{-1.8}^{+1.9} \ {\rm M_{Jup}}$. Case~(iv): like (iii), but increasing the pebble isolation mass to 100~$\rm M_\oplus$. The reasoning for testing these different cases will be discussed below, where we summarize the inversion results obtained for the different cases.

\begin{figure*}[t!]
\centering
\includegraphics[width=0.495\textwidth]{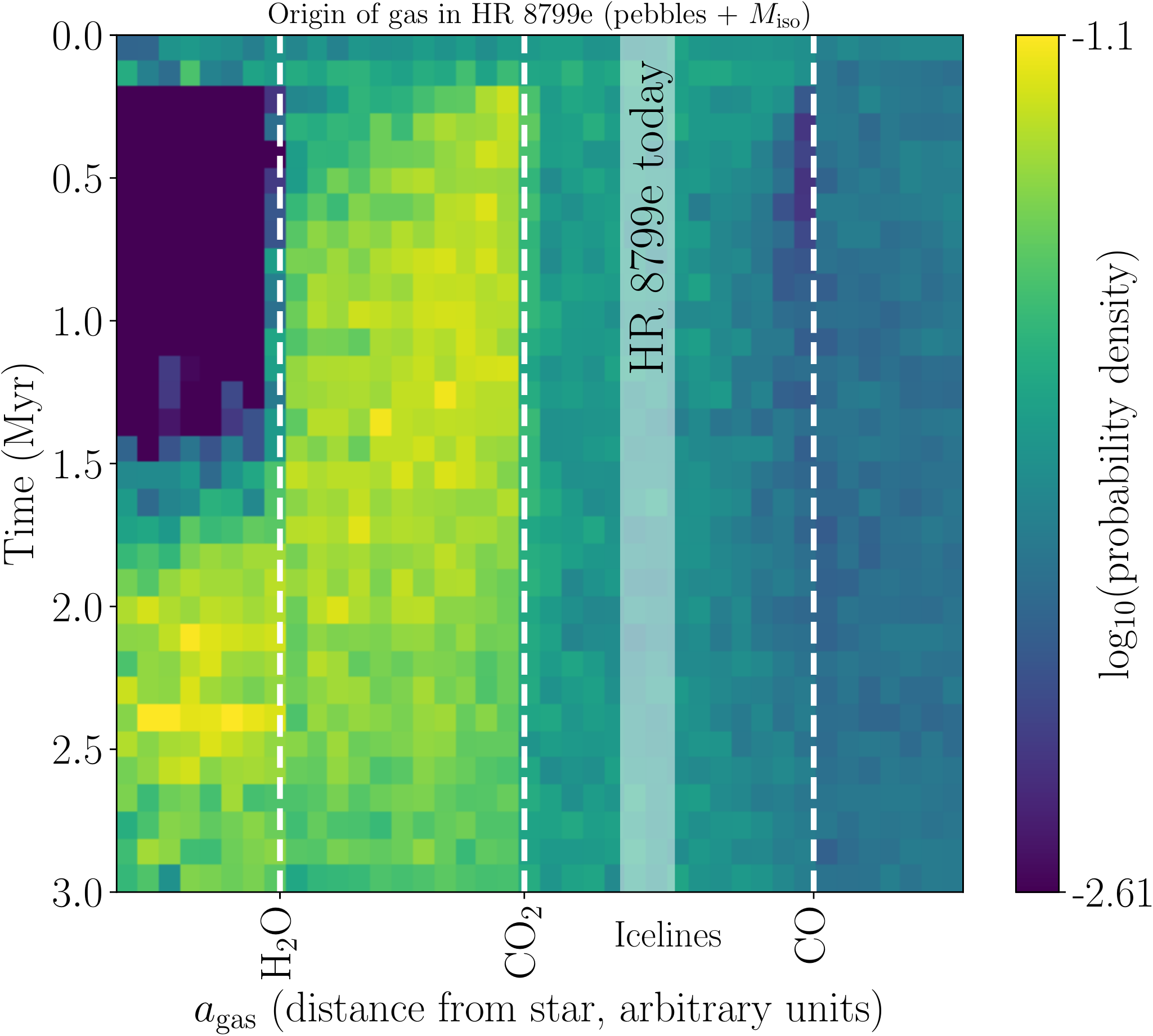}
\includegraphics[width=0.495\textwidth]{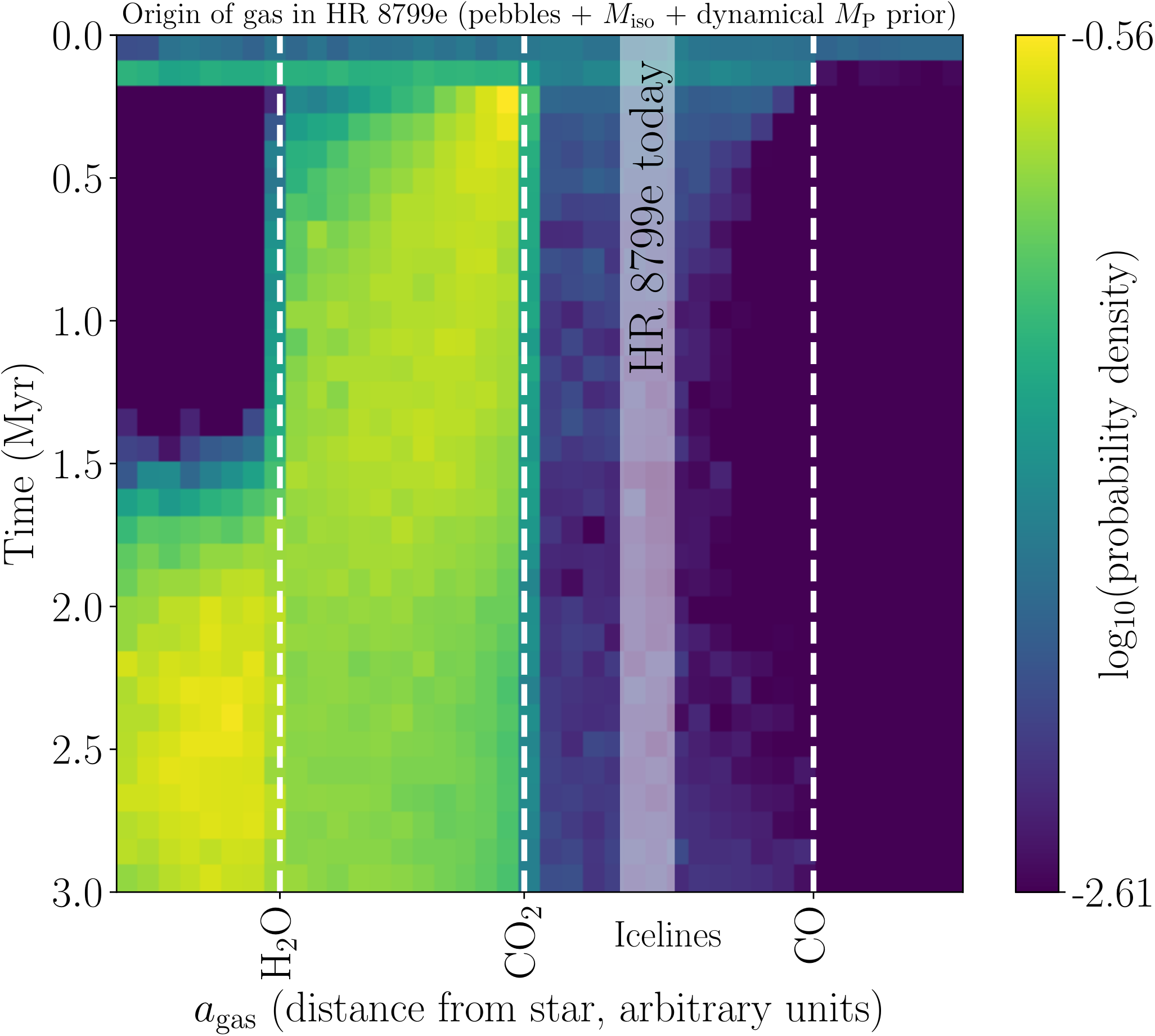}
\caption{\rrp{Time-dependent posterior of $a_{\rm gas}$, the location where gas was accreted by the forming planet HR~8799e. The iceline positions of \ce{NH3} and \ce{N2} have been omitted for clarity, and the x-axis has been log-scaled. {\it Left panel:} pebble drift scenario when including a pebble isolation mass prior of 20~$\rm M_\oplus$ for the accreted solids (Case ii). {\it Right panel:} like the left panel, but for Case (iii), that is additionally including a dynamical mass prior on the mass of HR~8799e.}}
\label{fig:gas_orig_HR8799e_pebble}
\end{figure*}

The compositional fit for Case (i) is depicted by the leftmost ellipse in the lower panel of Figure~\ref{fig:CoHOoHs_GRAVITY}. In this scenario pebble drift, evaporation, and accretion is able to reproduce the observed abundance pattern of HR~8799e. Conceptually, Case~(i) simply tests whether the results of the static disk model inversion change when introducing pebbles, but does not yet apply any prior knowledge on how pebbles are accreted, such as the concept of the pebble isolation mass. The reason for the good compositional fit of Case~(i) becomes evident when studying the right panel of Figure~\ref{fig:solid_orig_HR8799e_pebble}, which shows the resulting posterior for the most likely accretion location of the solids for HR~8799e. Because the solid C/O values do not change significantly, except for just outside the icelines, the result that significant accretion of solids from outside the CO iceline is likely does not change when compared to the static setup of the disk composition. Just outside the CO iceline the probability goes down, however, because CO gas recondensing on the pebbles drives up the pebbles' C/O to values larger than the planetary one. What is noticeable is that the region inside the CO iceline at $t>0$ is somewhat more likely when compared to $t=0$. This is because the disk gas, enriched by CO from evaporating pebbles, and with ${\rm C/O} = 1$, is of high enough metallicity to somewhat offset the C/O value of solids accreted inside the CO iceline, which is too low when compared to the planet. The enrichment of the disk gas in CO over time is shown in Figure \ref{fig:CO_increase}. We note that the likelihood for accreting a significant mount of pebbles decreases over time, because pebbles will drain to the inner parts of the disk. We neglect this effect here.

In Case (i) the upper limit on the planetary mass from the spectroscopic retrieval, together with a super-stellar atmospheric metallicity, leads to a 1-$\sigma$ upper limit of 570~$\rm M_\oplus$ on the accreted pebble mass. Such a high value is inconsistent with the concept of the pebble isolation mass. Therefore we deem Case (ii), where we set an upper limit of 20~$\rm M_\oplus$ on the accreted solid mass, a more likely scenario. We note that the pebble isolation mass is a function of the disk viscosity $\alpha$, and that it is very sensitive to the disk aspect ratio ($M_{\rm iso}\propto [H/r]^3$, with $H$ being the disk scale height). The value of 20~$\rm M_\oplus$ is what we derive for the HR~8799 disk model at the location of the CO iceline, using the scaling relations reported in \citet{bitschmorbidelli2018}. The compositional fit for Case (ii) is shown in the lower panel of Figure~\ref{fig:CoHOoHs_GRAVITY} (second ellipse from the left). Also in this scenario pebble drift is able to reproduce the observed abundance pattern of HR~8799e, but leads to a generally somewhat lower planetary metal enrichment. We note that these results assume that all accreted pebbles are visible in the atmosphere, which is equivalent to full core dissolution and mixing. Moreover, in order to allow pebble enrichment to have a noticeable effect on the planet composition, the inversion constrains the planetary mass to $<3.5 \ {\rm M_{Jup}}$. What is more, as a result of the prior limit on the accreted solid mass and the planetary mass prior, the inversion deems scenarios more likely where the composition of the accreted gas has more impact than in Case (i). The resulting probability distribution on the locations $a_{\rm gas}$ where the planet accreted its gas is shown in the left panel of Figure~\ref{fig:gas_orig_HR8799e_pebble}. The most likely locations and times for the gas accretion correspond to the situation where the gas enriched by the evaporating pebbles reaches approximately C/O values of 0.6, corresponding to the planet's atmosphere (cf. left panel of Figure~\ref{fig:solid_orig_HR8799e_pebble}). In this scenario the most likely formation accretion location is inside of HR~8799e's current orbital position, which would require some outward migration if taken at face value.

In Case (iii) we study the effect of enforcing an upper limit on solid accretion, due to the pebble isolation mass, and a tighter constraint on the planetary mass. The mass prior stems from a dynamical analysis based on the orbital characterization of the HR~8799 system, and accelerations from the \emph{Gaia-Hipparcos} catalog \citep{brandtbrandt2020}. The compositional fit for Case (iii) is shown in the lower panel of Figure~\ref{fig:CoHOoHs_GRAVITY} (second ellipse from the right). In this case the inversion struggles to reproduce the observed enrichment pattern of HR~8799e; while it fits the atmospheric C/O ratio well, the planetary enrichment is generally too low, but improves at later times. This is explained from the fact that the high mass prior assumed for HR~8799e, together with the low pebble isolation mass, does not allow for the pebbles to play a significant role in the planet enrichment, while especially at early times the disk gas is not enriched enough by gas that has evaporated off the inward drifting pebbles. This situation is thus alleviated at later times, when the disk gas enrichment increases, but it is never enough to fully reproduce the planetary metal enrichment. The right panel of Figure~\ref{fig:gas_orig_HR8799e_pebble} shows probability distribution of $a_{\rm gas}$. Because only gas accretion is able to affect the planetary composition noticeably in Case (iii), it is essentially a higher contrast version of the $a_{\rm gas}$ distribution of Case (ii), shown in the left panel.

\begin{figure}[t!]
\includegraphics[width=0.47\textwidth]{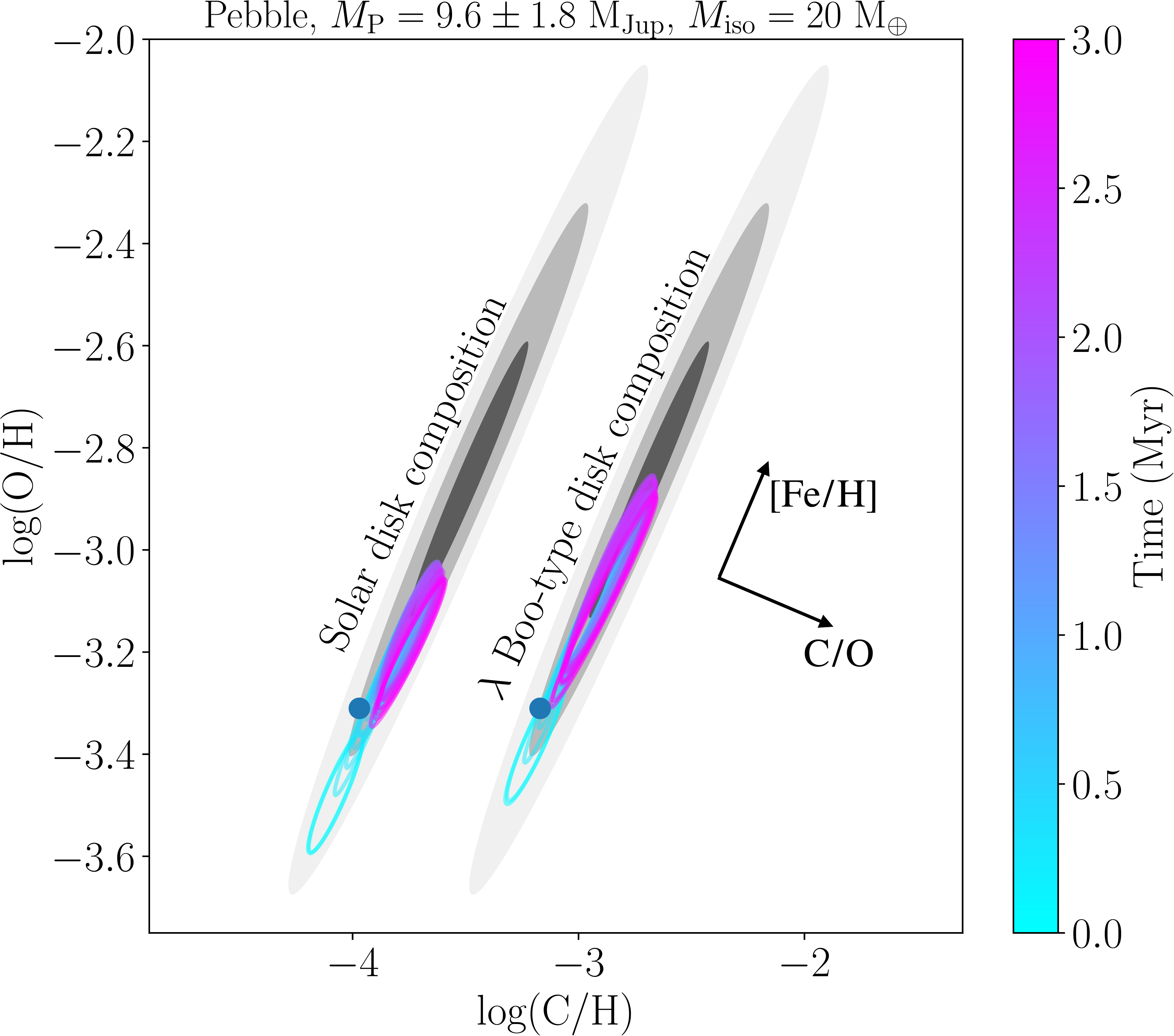}
\caption{\rrp{Like Figure~\ref{fig:CoHOoHs_GRAVITY}, but comparing the nominal pebble inversion Case (iii), that is $M_{\rm iso} =  20 \ {\rm M_\odot}$, $M_{P} = 9.6_{-1.8}^{+1.9} \ {\rm M_{Jup}}$, at solar disk composition (left ellipse) with a setup where the $\lambda$~Boo-type composition of HT~8799 was assumed for the disk instead (right ellipse). An offset was applied to these ellipses for clarity.}}
\label{fig:lambdaBooPebble}
\end{figure}

Case~(iv) essentially studies the case when the planet started forming very far outside the CO iceline, in the outer parts of the disk. Due to the disk flaring, $H/r$ increases towards the outer disk, and we would find $M_{\rm iso} = 50 \ {\rm M_{\oplus}}$, corresponding to $H/r=0.07$, at 200~au. Because we are interested in an upper limit on what pebble accretion could contribute we also assume that the disk viscosity is very high for the $M_{\rm iso}$ calculation ($\alpha = 0.004$, instead of the nominal 0.0005); this results in  $M_{\rm iso}=100 \ {\rm M_\oplus}$. The corresponding enrichment pattern of the planet is shown in the lower panel of Figure~\ref{fig:CoHOoHs_GRAVITY}, rightmost ellipse. Unsurprisingly, note only the planet C/O but also its metal enrichment are better fit now, when compared to Case~(iii), leading to a good fit overall. Like expected, $a_{\rm solid}$ values outside the CO iceline are the most likely for this case, with some additional gas accretion from within the \ce{CO2} iceline.

From our investigation it thus becomes evident that pebbles alone, for average $M_{\rm iso}$ values, may not be sufficient to fully explain the observed abundance pattern of HR~8799e, even when making the assumption that all pebbles accreted onto the planet (likely onto the forming planetary core) mix into the visible atmosphere. This conclusion hinges on at least three assumptions. First, if the planetary mass was actually lower than reported in \citet{brandtbrandt2020}, enriching the planet by the accreted solid pebbles becomes easier. This is shown by the our Case (ii), where the inversion when imposing $M_{\rm iso}=20$~$\rm M_{ Jup}$ resulted in an a good fit by constraining the planetary mass to below 3.5~$\rm M_{Jup}$. Next, if the pebble isolation mass is much higher than our baseline case (e.g., 100 instead of 20~$\rm M_{\oplus}$), which is possible for large disk viscosities and the planet initially forming far out in the disk, pebble enrichment becomes a likely scenario for explaining the abundance pattern of HR~8799 again. Lastly, if the composition of the disk is different from our baseline case, even $M_{\rm iso} = 20 \ \rm M_{\oplus}$ with the dynamical mass prior of HR~8799e (Case iii) becomes a likely scenario again. This is seen in Figure~\ref{fig:lambdaBooPebble}, where we show what happens when running Case (iii) again, but assuming the $\lambda$~Boo-type composition of HR~8799 for the disk composition. Due to the carbon and oxygen content being $\sim$30~\% higher in this case, the accretion of gas that is enriched by evaporated pebbles leads to a better agreement with the total atmospheric metal enrichment. We note that all of these conclusions are based on the spectroscopic retrieval result for HR~8799e, and a slightly lower retrieved metallicity would make pebble accretion more likely again. Due to the large uncertainties on the atmospheric metallicity, even the pebble scenario with the worst fit (Case iii) is only about one standard deviation away from the mean composition derived in the spectroscopic retrievals.

Lastly, it should be kept in mind that other likely important effects connected to pebbles were not studied here. For example, we neglected the effects that the outer HR~8799 planets may have had on the pebble flux that reaches the inner disk, and therefore HR~8799e's position. Outer planets may prevent pebbles from drifting inward and evaporating at the CO iceline \citep[e.g.,][]{bitschraymond2021,schneiderbitsch2021}. It is unclear to what degree this effect is important here, because the giant planets may have formed late enough that some pebble drift may already have taken place in the disk before shutting off the pebble flux. In addition, if HR~8799e, the innermost planet, formed first (high surface densities and orbital periods, thus shorter accretion timescales) it may have been less affected by the formation of the outer planets.}

\subsection{Suggested toy models to study other formation aspects}
\label{subsect:toy_model_inversions}

Above it was studied how inferences drawn from a simple formation model change if chemical evolution of the protoplanetary disk, \rrp{or the drift, evaporation and accretion of pebbles} is included. As discussed in Section~\ref{sect:planet_form_complexity}, planet formation is the combination of quite a number of key processes. A concurrent formation inversion with all the ingredients appears both numerically and conceptually unworkable, at the moment. It will still be instructive, however, to add certain aspects of the planet formation problem to such inversion calculations, to study their influence in isolation, or to assess the magnitude of their importance for atmospheric compositions. In Table~\ref{tab:formation_challenges_in_inversions} we list the way in which many of the aspects mentioned in Section~\ref{sect:planet_form_complexity} may be studied via inversion of the formation process.

\begin{table*}[t!]
\centering
{ \footnotesize
\begin{tabular}{ll}
\hline \hline
Aspect & Potential tractability in formation model inversions \\ \hline \hline
{\it Disk composition and structure}  &  \\ \hline 
Unknown disk elemental abundances  & Scale using stellar [Fe/H], try varying composition according to scaling uncertainties. \\ 
Available solid reservoir  & Impose limit based on likely disk mass and dust-to-gas ratio. \\ 
Disk (thermal) structure  & Feed in disk structures from dedicated disk models. \\
& Explore if parameterizing 3-d effects in 1-d model is possible.\\
& Changes in disk structure will affect, e.g., iceline positions, as in \citet[][]{ohnoueda2021}. \\
Planetary back-reaction on disk  & Use simplified gap opening criteria to limit gas accretion, compare to disk lifetimes. \\
& Apply pebble isolation mass (limit refractory reservoir accessible to planet). \\
Include pebble drift \& evaporation at icelines  & Increase gas metallicity inside of icelines as function of time, also see Sect.~\ref{subsect:formation_inversion_hr8799e}. \\ \hline \hline
{\it Disk chemistry} & \\ \hline
Chemical evolution of disk  & Run formation inversion with chemical composition as function of time. \\
& Also see Section \ref{subsect:formation_inversion_hr8799e}. \\
Inherited or `reset' disk abundances  & Explore impact of differing assumptions on disk abundances for the inversion process. \\
Cosmic ray ionization and stellar irradiation  & Use best guesses for retrievals, otherwise explore different values. \\
Refratory carbon depletion in inner disk  & Explore via on/off switch. \\ \hline \hline
{\it Planet formation}  & \\
\hline
Pebble and planetesimal accretion  & Compare inferred solid (refractory) mass of planet with isolation masses.  \\
& Constrain upper limit on accreted $M_{\rm pebbles}$, lower limit on accreted $M_{\rm planetesimals}$. \\
3-d planet accretion & Test impact of parameterizations, for example vertically averaged \\
& abundances for gas accretion \\
Planet migration  & Allow to fit for multiple formation locations? \\
& Add priors on formation location: planet traps? \\
& Add priors enforcing inward migration (e.g., $a_{\rm gas} \leq a_{\rm solid}$)? \\
Leveraging full complexity of formation models  & Explore use of machine learning techniques. \\
& E.g., random forest predictors as demonstrated in \citet{schleckerpham2021}. \\
Planet formation by gravitational instability  & Likely treatable, but requires changes. \\
& E.g., steady state viscous disk model $\rightarrow$ infall disk model\\ \hline \hline 
{\it Planet bulk -- atmosphere coupling}  & \\ \hline
Metallicity gradient inside planet  & Use multi-component model, infer mixing efficiency $f_{\rm mix} \in [0,1]$ \\
& to reveal correlations with other formation parameters. \\
\hline \hline
{\it Atmospheric evolution} & \\ \hline
Evaporation  & Can be important for lower mass planets or gas planets with a metallicity gradient. \\
& Use inverted evaporation models to reveal correlation with formation parameters? \\
Infall of comets / asteroids  & Better quantitative modeling needed. \\
& Potentially not important for gas giant planets. \\ \hline 
\end{tabular}
}
\caption{Aspects of planet formation and their potential treatment toy formation model inversions.}
\label{tab:formation_challenges_in_inversions}
\end{table*}

\rrp{To give an example, it would be straightforward to feed disk compositional models that include the disk's self-shadowing into the inversion framework. This process has been suggested by \citet{ohnoueda2021}, where the shadowing is caused by a dust pile-up at the water iceline. Depending on the grain properties and densities, such a scenario may allow for very volatile species such as CO, \ce{N2} and even noble gases such as Ar, to condense at distances from the star that are nominally too hot. To study such an effect, various mid-plane disk and abundance structures, for differing dust density contrasts, could be explored.}

\rrp{Another setup that would be instructive is to further investigate the effect of incomplete mixing between the deep interior and planetary atmosphere for gas giant planets.} As discussed in Section~\ref{sect:planet_form_complexity}, the metal enrichment inferred from atmospheric characterization studies is only a lower limit for the true planetary metal enrichment. \rrp{Where available, an upper limit could be placed based on the analyses of planetary bulk metallicities, as obtained in \citet[][]{thorngrenfortney2016,thorngrenfortney2019b}.} The impact of metallicity gradients could potentially be studied by adding a parameter $f_{\rm mix}$ which describes whether the metals accreted during formation fully mix ($f_{\rm mix}=1$) into the atmosphere, or not ($f_{\rm mix}=0$). As long as a planetary atmosphere is of super-stellar metallicity, $f_{\rm mix}$ will simply be inversely correlated with the accreted solid mass (if pebble evaporation is neglected). Once a planetary atmosphere is of stellar or substellar metallicity, $f_{\rm mix}$ may also correlate with the formation location of a planet, depending on the disk's abundance structure. \rrp{An example for this can be constructed by considering our inversion results for HR~8799e in the static disk picture. Because the atmospheric metallicity is high, and the planet has a stellar C/O value, the inferred atmospheric C/O ratio could only be reproduced by accreting solids from outside the CO iceline. If the planet's atmospheric metallicity was stellar, it could instead have formed at any location in the disk, as long as the location of gas accretion is equal to the location of solid accretion (measured with respect to the icelines). If $f_{\rm mix}$ was added as a free parameter, small $f_{\rm mix}$ values would again have yielded regions outside the CO iceline as the most likely region of origin for the solids accreted by the planet.}

It is also conceivable to construct a three-component model for the formation inversion which separates the planetary mass into three reservoirs: solids accreted onto the core, solids accreted and mixed into the gaseous envelope, and the gas itself. Each of these would also be associated with a parameter that described where the corresponding material was accreted. One could then define an $f_{\rm mix}$ for the core (or solids in the deep interior) which would describe the degree to which the deep core dissolves and mixes into the envelope.

\rrp{Lastly, abundance constraints on the refractory content of a planet, as presented in \citet[][]{lothringerrustamkulov2020} for WASP-121b, for example, may be used to put an upper limit on the mass a planet accreted through pebbles. As long as the planet did not form very close around its star, where also refractories may enter the gas phase, refractories can only be incorporated into the planet by solid accretion. If the inferred amount of accreted refractories is higher than expected from the concept of the pebble isolation mass (see Section~\ref{sect:planet_form_complexity}), an upper limit on the amount of accreted pebbles, as well as a lower limit on the amount of accreted planetesimals (or other impactors such as smaller planets, e.g., \citealt{ginzburgchiang2020}), may be constrainable. Similar constraints may be obtained from the cases where the volatile enrichment of a planet is higher even than what pebble evaporation in a disk may provide: any additional volatile mass must then be accreted in the form of volatile ices.}
\section{Future observatories and a census of atmospheric compositions}
\label{sect:obs_future}

In the previous sections we discussed that inverting atmospheric compositions to reveal the detailed formation history of a planet is hardly at the moment: the process of planet formation is too complex, with too many unknowns, and likely too numerically costly to invert. However, the {\it James Webb Space Telescope}, the class of future ground based Extremely Large Telescopes (ELTs), and later {\it ARIEL} will record high-quality spectra for hundreds of planets. In this section we summarize the compositional constraints that can be extracted from such atmospheric measurements and how the resulting atmospheric enrichment patterns for the planetary population may allow to constrain planet formation in a broader sense.

\begin{figure*}[t!]
\centering
\includegraphics[width=0.98\textwidth]{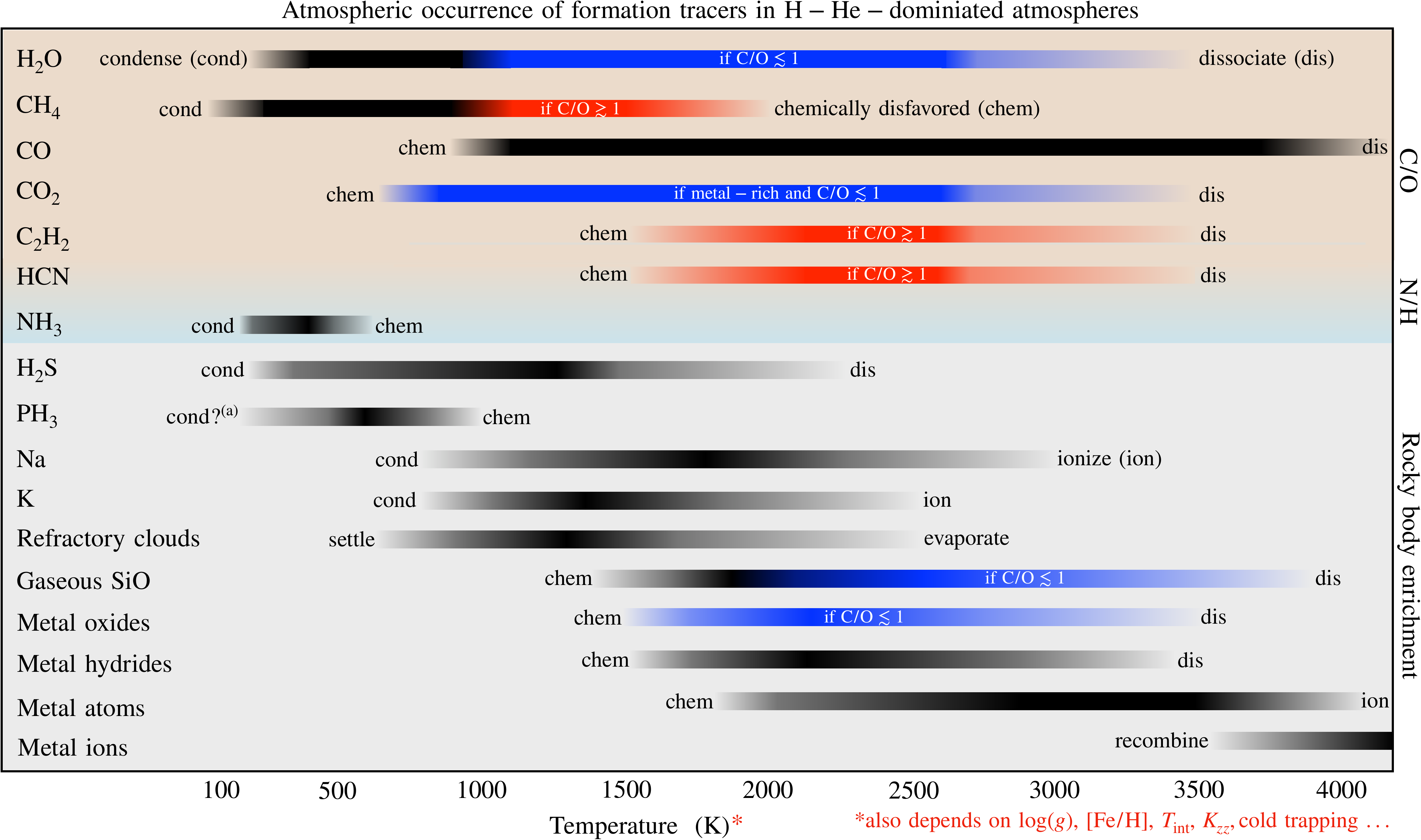}
\caption{Potential atmospheric visibility of various absorbers in planetary atmospheres. Every species or group of species shown here is known to be spectrally active. We searched the literature for the average atmospheric temperatures where these species are visible. Alternatively we used the equilibrium chemistry code described in \citet{mollierevanboekel2016} and checked for which temperature range the species is present in the atmosphere. Our standard assumption was solar metallicity and abundance ratios, and a pressure of 0.1 bar, whereas dissociation and ionization values were obtained from assuming pressures from 0.1 to 0.001 bar. We either assumed solar C/O ($=0.55$) or ${\rm C/O}=1.1$. The temperatures given therefore should only serve as rough guidance, and do not necessarily correspond to a planet's effective temperature. We note that chemical transitions also depend on the metallicity, and the pressure at the planetary photosphere (therefore effectively also on the planetary gravity $g$). Moreover, many of these species can be affected by disequilibrium chemistry \citep[see, e.g.,][]{fortneyvisscher2020}, or be cold trapped into condensates \citep[e.g.,][]{spiegel2009,parmentierfortney2016}. The chemical behavior of the species listed here is described in Section \ref{sect:obs_future}, and Appendix \ref{sect:refractory_chem} for the refractories.}
\label{fig:atmo_detect_composition}
\end{figure*}

\subsection{C/O}
\label{sect:form_tracers:ctoo}

\rrp{The importance of the planetary C/O ratio for informing planet formation has been discussed in Sections~\ref{sect:introduction} and \ref{sect:planet_form_complexity}. In these sections we also discuss the complications that may make the picture likely more complex than suggested by the foundational study by \citet{oeberg2011}. For example, while dominant solid enrichment is generally expected to lead to stellar or sub-stellar C/O ratios and super-stellar planetary metallicities, pebble drift and evaporation may lead to super-stellar enrichments of the gas at C/O values both smaller or larger than stellar. Interestingly, however, super-stellar C/O values and enrichments are difficult to obtain without considering pebble evaporation (also see Section~\ref{subsect:formation_inversion_hr8799e}), so a large population of planets which such abundance characteristics may indicate a dominant role of pebbles for setting planetary abundances. Similarly, a large enough overall metal enrichment of a planet, especially if formed in the outer disk, may be difficult to explain from pebble evaporation, even for low disk viscosities, which would point more towards planetesimal accretion playing an important role.

In general, C/O is also popular because it determines the relative abundances of the spectrally active C- and O-bearing molecules in the exoplanet atmospheres such as \ce{H2O}, \ce{CH4}, CO, \ce{CO2}, HCN, \ce{C2H2}. C/O therefore regulates the spectral appearance of a planet in the near- to mid-infrared  \citep[e.g.,][]{seager2005,fortneymarley2005,madhusudhan2012,mosesmadhusudhan2013,mollierevanboekel2015,molaverdikhanihenning2019,goyalmayne2020,hobbsmadhusudhan2021}.
 
For reference, Figure \ref{fig:atmo_detect_composition} shows under which atmospheric conditions the absorbing species that trace the C/O ratio in gas-dominated planets may be visible. Also see, for example, \citet[][]{loddersfegley2002} for a detailed description of the atmospheric chemistry. For temperatures below about 1000~K the atmosphere will be rich in \ce{H2O} and \ce{CH4}, for higher temperatures these species will be converted into \ce{CO} until either C or O runs out, depending on the C/O ratio. For high temperatures and ${\rm C/O}\gtrsim 1$, \ce{CH4} will thus be visible; for high temperatures and ${\rm C/O}\lesssim 1$, \ce{H2O} will be visible. For further increasing temperatures and ${\rm C/O}\gtrsim 1$, \ce{CH4} is replaced by increasing amounts of \ce{C2H2} and \ce{HCN} \citep[see, e.g.,][]{madhusudhan2012,mollierevanboekel2015}. We note that the chemical transitions mentioned here also depend on the local atmospheric pressure \citep[][]{mollierevanboekel2015,molaverdikhanihenning2019}.  \ce{CO} can still be visible in cool atmospheres, especially of self-luminous brown dwarfs and planets, because atmospheric mixing may transport CO-rich gas from the deep (hotter) atmosphere to the photosphere \citep[e.g.,][]{zahnlemarley2014,milesskemer2020}. Whether such disequilibrium abundances are expected for irradiated (often transiting) planets is less clear, because the insolation leads to more isothermal atmospheres. For planets which are still strongly cooling (with a high internal temperature) or heated by processes such as eccentricity dampening, \ce{CH4} may be strongly suppressed \citep{fortneyvisscher2020}. As mentined, ground-based high-contrast or high-resolution observations have started to obtain the first useful constraints on C/O. The state-of-the-art will greatly improve once \emph{JWST} and later \emph{Ariel} will allow for a larger census of planetary compositions (also see our discussion in section~\ref{sect:introduction}).}
 
\subsection{N/O, N/C}
\label{sect:nitrogen_content}
The importance of atmospheric nitrogen-bearing species such as \ce{NH3} or \ce{HCN} for constraining exoplanet formation has been recognized recently, especially for planets forming in the outer solar, and extrasolar disks. The reason for this is that nitrogen, predominantly in the form of \ce{N2} in protoplanetary disks, is extremely volatile. Planets forming at increasingly larger distances, when dominated by solid metal enrichment, will therefore exhibit increasingly lower N/O or N/C ratios, and vice versa if dominated by gas metal enrichment. This is because several icelines of C and O-bearing species are crossed towards larger orbital radii, while \ce{N2} stays in the gas phase \citep{turrinischisano2020}. If the planet forms at wide enough orbital distances, eventually \ce{N2} will freeze out as well, leading to an enhanced atmospheric nitrogen content, which will scale similarly with atmospheric metallicity as the abundances of C- and O-bearing species. The high nitrogen content of Jupiter has therefore led to the interpretation that Jupiter formed in the outer regions of the Solar System, beyond the location of the \ce{N2} iceline at $\sim 30$~au, which is also consistent with the planet's elevated abundance of noble gases \citep[e.g.,][]{owenencrenaz2003,bitschlambrechts2015,oebergwordsworth2019,bosmancridland2019,cridlanddishoek2020}. Similar to the the discussion of C/O, the situation is likely more complicated also for the nitrogen enrichment. Both disk self-shadowing \citep[see][and our discussion in Section \ref{subsect:toy_model_inversions}]{ohnoueda2021} or pebble drift and evaporation \citep{schneiderbitsch2021b} are likely complicating factors.
We also note that a planet that forms late within a protoplanetary disk's lifetime may be less sensitive to \ce{N2}, as cosmic ray ionization may process \ce{N2} to \ce{NH3} ice over Myr timescales, such that the importance of \ce{NH3} and its iceline increases over time, with the iceline of \ce{NH3} being much closer to the star than the one of \ce{N2} \citep{2011ApJS..196...25S}.

In exoplanets the only spectrally active nitrogen bearing species of relevance are \ce{NH3} and HCN. \ce{N2}, which is the dominating nitrogen bearer at larger temperatures, has  negligible opacity in the near- and mid-infrared. \ce{NH3}, on the other hand, should be detectable in the mid-IR using {\it JWST} in exoplanet atmospheres \citep[e.g.,][]{danielskibaudino2018}. \rrp{Moreover, evidence for \ce{NH3} has been seen in high-resolution studies \citep[][]{giaccobebrogi2021,sanchezlopezlandman2022}}. For $\rm C/O \lesssim 1$, \ce{NH3} is only abundant up to $\sim 500$~K \citep{loddersfegley2002}. HCN, on the other hand, will be visible for temperatures of 1500~K or larger, if ${\rm C/O}>1$ \citep[e.g.,][]{madhusudhan2012,mollierevanboekel2015}. We indicate these detectability ranges in Figure \ref{fig:atmo_detect_composition}. Chemical disequilibrium may (or may not) allow for \ce{NH3} or HCN to be visible at intermediate temperatures ($500 \ {\rm K} < T < 1500 \ {\rm K}$) in irradiated planets as well \citep[see, e.g.,][and the references therein]{macdonalmadhusudhan2017b,fortneyvisscher2020,hobbsrimmer2021}. For self-luminous planets disequilibrium chemistry may play less of a role for N, as iso-abundance lines are parallel to atmospheric pressure-temperature profiles \citep{zahnlemarley2014}. \rrp{As before, all chemical transition temperatures also depend on the atmospheric pressure.}

We note that the chemical behavior of N, C and O bearing species described here mostly hinges on chemical equilibrium or simple atmospheric disequilibrium treatments, also considering the planetary atmospheres to be one-dimensional and of mostly scaled solar abundances (except for the C/O ratio). The recent and intriguing results of \citet{giaccobebrogi2021}, who detected \ce{H2O}, CO, \ce{HCN}, \ce{C2H2}, \ce{NH3}, and \ce{CH4} in the atmosphere of HD~209458b (with an equilibrium temperature of $\sim 1500$~K) are a reminder that atmospheric chemistry may be much more complex than discussed above. An important effect is the horizontal advection of chemical abundances predicted from coupling chemical models to the output of 3-d general circulation models \citep[e.g.,][]{agundezparmentier2014,baeyensdecin2021}, \rrp{which could also be connected to condensate rain-out \citep[][]{sanchezlopezlandman2022}}. Also photochemistry is important, especially in the upper atmospheric layers \citep[e.g.,][]{venot2012,kopparapu2012,molaverdikhanihenning2019b}. Telescopes such as JWST, current high-resolution spectrographs, and ultimately instruments mounted on ELT-class telescopes will allow us to investigate these effects more thoroughly.

\subsection{R/O}
Measuring the refractory content of an atmosphere could provide unique insight into a planet's formation history. As has been argued recently by \citet{lothringerrustamkulov2020}, measuring the refractory-to-oxygen ratio R/O of a planet constrains the importance of metal enrichment by rocky accretion relative to icy or gaseous accretion. Here R stands for any element that traces the refractory content of the planet (Fe, Na, K, Si, Mg, Ti, ...), or an average of such elements. As argued further, this may allow the placement of constraints on whether the planet (or its solid building blocks) migrated significantly during formation. \rrp{We argue that R/O may potentially even be useful to constrain the relative importance of pebble and planetesimal accretion}, in the core accretion paradigm, \rrp{also see Section \ref{subsect:toy_model_inversions} and the discussion in \citet[][]{schneiderbitsch2021b}.}

In Figure \ref{fig:atmo_detect_composition} we indicate the temperature ranges over which various refractory-tracing atmospheric absorbers are visible. We refer the reader to Appendix \ref{sect:refractory_chem} for a discussion of the chemistry of the refractory-tracing absorbers.  \citet{lothringerrustamkulov2020} put emphasis on ultra-hot Jupiters, for which various refractory elements exist as molecules (metal oxides or hydrides), atoms, or ions in the gas phase. We also note that species such as \ce{H2S} and \ce{PH3} may be useful refractory tracers at intermediate atmospheric temperatures \citep{wangmiguel2017,oebergwordsworth2019}. While  \ce{H2S} and \ce{PH3} are volatile species, the dominant carrier of P and S atoms in a protoplanetary disk appear to be refractory species \citep{oebergwordsworth2019}. Moreover, measuring the abundances of Na and K in planetary atmospheres may be worthwhile tracers of the refractory content \citep{welbanksmadhusudhan2019}.

Refractory cloud species may affect planetary spectra by muting molecular features and reddening the spectral energy distribution. Silicate particles like \ce{MgSiO3} and \ce{Mg2SiO4} are especially interesting, as they may lead to visible absorption features around 10~micron \citep[e.g.,][]{cushingroellig2006,wakefordsing2015}. Due to the complex micro-physical problem of cloud formation \citep[e.g.,][]{rossow1978,powellzhang2018,woitkehelling2020}, measuring a refractory abundance from observed cloud absorption may prove difficult, however. Moreover, clouds may complicate measuring and interpreting the abundances of gas refractory species due to cold trapping by condensate rainout \citep[e.g.,][]{spiegel2009,parmentierfortney2016}. An interesting alternative to silicate clouds could be searching for the absorption of gaseous SiO at $\sim$7~micron with JWST's MIRI instrument. SiO is promising because it is the most abundant Si-bearing gas species after the silicates evaporate \citep{visscherlodders2010}, and is more stable than \ce{H2O} against dissociation (by about 500~K). This should allow detecting this species in ultra hot Jupiters.

Other metal oxides such as TiO, VO, AlO, CaO have features in the optical and near-infrared \citep[e.g.,][]{sharpburrows2007,gandhimadhusudhan2019,lothringerfu2020}. Similarly, metal hydrides may be useful refractory tracers, at similar temperatures as the metal oxides. Species such as FeH, CaH, MgH, NaH, CrH, TiH all have absorption features in the optical and near-infrared \citep{sharpburrows2007,gandhimadhusudhan2019}. Metal atoms are visible in the atmosphere once the refractory clouds are no longer present (e.g. Mg, Fe), or once the dominant molecular species (such as SiO for Si) have been dissociated. Mg, Fe, Ca, Cr, Ni, V, Na and maybe Co have been detected in the ultra hot atmospheres of KELT-9b and WASP-121b in the optical \citep{hoeijmakersehrenreich2019,hoeijmakersseidel2020}. Finally, metal ions become visible in the hottest atmospheres as soon as the atoms have been ionized. This has led to the detection of Fe+, Ti+, Cr+, Sc+, Y+ and maybe Sr+ in the hottest known exoplanet KELT-9b in the optical \citep{hoeijmakersehrenreich2019}.
\section{Discussion and summary}
\label{sect:summary}

Inferring the formation history of a planet, based on its atmospheric composition, is one of the most cited goals of the atmospheric characterization community. In our work we take a look at what obstacles need to be overcome to make such an inversion feasible.

Summarizing the complex and interconnected processes that govern planet formation (see Section~\ref{sect:planet_form_complexity}), we conclude that actually inverting planet formation in this way is still a long way off, if even possible at all. Current formation models are likely too complex (too many free parameters), too uncertain (which processes to consider, which assumptions to make for them), and too numerically costly (N body interactions, dust evolution, hydrodynamical evolution, disk chemical evolution, etc.). Many of these problems may actually be alleviated in the coming years or decades, but the degree to which such a full formation inversion will ever become possible is difficult to assess, at the moment. As an interesting avenue for inverting full, state-of-the art formation models, we want to highlight the recent work by \citet{schleckerpham2021}, where a random forest technique was used to predict planetary formation outcomes based on formation model input parameters. It will have to be seen in how far this method can be used to predict planetary abundances.

Apart from this conclusion, we also introduce a method that allows to study and compare the qualitative impact of different assumptions made in the modeling of planet formation, see Section \ref{sect:formation_inversion}. Assuming some measured planetary compositions as observations, we use nested sampling to invert simplified formation models, constraining their corresponding formation parameters. Due to the challenges mentioned above, such invertible formation models cannot be complex enough to yield reliable results on a given planet's formation process. However, they may allow to study the importance of various formation aspects in isolation. As an example, we show how the deduced formation history of the directly imaged planet HR~8799e changes if the composition of the protoplanetary disk in which it forms is allowed to evolve chemically. We find that chemical evolution may significantly affect the migration history inferred for this planet; the planet may have migrated much less if chemical evolution is taken into account. \rrp{What is more, we show that the drift, evaporation, and accretion of pebbles is able to reproduce the planetary C/O value, but whether it can reproduce the inferred high atmospheric metallicity depends on the  assumptions made for the disk viscosity, pebble isolation mass and the disk composition.} We end this section by suggesting a number of other formation processes that could be studied in a similar way, for example metallicity gradients and ineffective mixing of the planetary interior.

While the detailed inversion of planet formation may still be in the far future, it is clear the atmospheric abundance constraints obtained with new and upcoming instruments will be crucial to inform planet formation models in a broader sense. In Section \ref{sect:obs_future} we summarize under which atmospheric conditions various spectrally active atmospheric species that trace the C/O value (\ce{H2O}, \ce{CO}, \ce{CH4}, \ce{CO2}, \ce{C2H2}, \ce{HCN}), nitrogen content (\ce{NH3} and HCN), and refractory content (\ce{H2S}, \ce{PH3}, alkalis, refractory clouds, metal oxides, hydrides, atoms and ions) may be observable in H/He-dominated atmospheres. Instruments such as {\it GRAVITY}, {\it CRIRES+} (or other high-resolution spectrographs), {\it JWST}, and facilities further in the future like {\it ARIEL} and the ELTs will obtain abundance constraints for many of these species. We discuss how the C/O values derived for the atmospheric composition of exoplanets may allow us to constrain the importance of pebble drift and evaporation, and how the refractory content of a planet may constrain the relative contribution of planetesimal and pebble accretion.

Making the connection between atmospheric abundances and formation a reality seems daunting, but the likely transformative nature of observations of many upcoming observational facilities will lead to more precise atmospheric abundance constraints for exoplanets than ever before. The constraints obtained from these observations will require being put into context, to assess what information on planet formation may possibly be gleaned from them. With these data one may begin assessing the degree to which planet formation can indeed be informed by the atmospheric composition of exoplanets.

\begin{acknowledgments}
We would like to thank the anonymous referee for their detailed report, which greatly improved the quality of this manuscript. P.M. and Th.H. acknowledge support from the European Research Council under the European Union's Horizon 2020 research and innovation program under grant agreement No. 832428-Origins. T.M. acknowledges support of Ministry of Science and Higher Education of the Russian Federation under the grant 075-15-2020-780 (N13.1902.21.0039; Section 2.2 \& 3.1.2). B.B., thanks the European Research Council (ERC Starting Grant 757448-PAMDORA) for their financial support. A.S. acknowledges funding from the European Union H2020-MSCA-ITN-2019 under grant agreement no. 860470 (CHAMELEON). R.B. acknowledges the support from the Swiss National Science Foundation under grant P2BEP2\_195285. C.M. acknowledges the support from the Swiss National Science Foundation under grant BSSGI0\_155816 ``PlanetsInTime''. D.S. acknowledges support by the Deutsche Forschungsgemeinschaft through SPP 1833: ``Building a Habitable Earth'' (SE 1962/6-1). Parts of this work have been carried out within the frame of the National Center for Competence in Research PlanetS supported by the SNSF.
\end{acknowledgments}

\appendix

\section{Atmospheric evolution}
\label{appendix:atmospheric_evolution}

Here we derive Equation \ref{equ:deltaX}, which estimates the change in mass fraction of a given species due to the enrichment of a planet's atmosphere by impacts, also see Section \ref{subsect:atmospheric_evolution}. We also discuss the case of pure water comets increasing the atmospheric water content of a Jovian planet.

To begin we estimate the mixing in the atmosphere by 1-d diffusion.
We thus write, using the usual 1-d diffusion equation for concentrations \citep[e.g.,][]{parmentiershowman2013}: 
\beq
\rho \frac{\partial X}{\partial t} = \frac{\partial}{\partial z}\left(K_{zz}\rho\frac{\partial X}{\partial z}\right) + W(z,z_{\rm i}) \frac{\dot{M}}{4\pi R_{\rm P}^2\Delta z_{\rm i}},
\label{equ:diff_alt}
\eeq
where $\rho$ is the atmospheric density, $X$ the atmospheric water mass fraction, $t$ the time, $z$ the atmospheric altitude, $\dot{M}$ the mass accretion rate of pure-water comets, and $R_{\rm P}$ the planetary radius.
$W(z,z_{\rm i})$ is defined as
\beq
W(z,z_{\rm i}) = \Theta(z-z_{\rm i}+\Delta z_{\rm i}/2) - \Theta(z-z_{\rm i}-\Delta z_{\rm i}/2),
\eeq
with $\Theta$ being the Heaviside step function. This means that we assume that the comets are destroyed in a narrow layer of width $\Delta z_{\rm i}$, at altitude $z_{\rm i}$ in the atmosphere. Using the equation of hydrostatic equilibrium ($\partial P / \partial z = - \rho g$), together with the equation of state of an ideal gas ($P=\rho k_{\rm B} T / \mu$) and that $H_P = k_{\rm B}T/\mu g$ one finds that once can express Equation \ref{equ:diff_alt} as
\beq
\frac{\partial X}{\partial t} = \frac{\partial}{\partial P}\left(\frac{K_{zz}}{H _P^2}P^2\frac{\partial X}{\partial P}\right) + W(P,P_{\rm i})\frac{\dot{M}g}{4\pi R_{\rm P}^2\Delta P_{\rm i}},
\label{equ:time_evo_X}
\eeq
where $P$ is the atmospheric pressure, $g$ the gravity, $k_{\rm B}$ the Boltzmann constant, $T$ the atmospheric temperature and $\mu$ the atmospheric mean molecular weight.
For a steady state ansatz and integration over $P$ one obtains that
\beq
\frac{\partial X}{\partial P} = - \frac{\dot{M}g}{4\pi R_{\rm P}^2}\frac{H _P^2}{K_{zz}}\frac{1}{P^2}
\label{equ:dXdP}
\eeq
for $P\geq P_{\rm i}+\Delta P_{\rm i}/2$, and $\partial X / \partial P = 0$ for $P\leq P_{\rm i}-\Delta P_{\rm i}/2$, and a linear transition between these two cases for $P \in (P_{\rm i}-\Delta P_{\rm i}/2,P_{\rm i}+\Delta P_{\rm i}/2)$. In the following we assume that $\Delta P_{i\rm } \ll P_{\rm i}$. Thus it holds that the enrichment in units of mass fractions $\Delta X = X - X_0$ at $P_{\rm i}$ and at lower pressures (higher altitudes) is
\beq
\Delta X = \frac{\dot{M}g}{4\pi R_{\rm P}^2}\frac{H _P^2}{K_{zz}}\frac{P_{\rm RCB}-P_{\rm i}}{P_{\rm RCB}P_{\rm i}},
\label{equ:deltaXApp}
\eeq
where $K_{zz}$ was assumed to be constant for simplicity. $P_{\rm RCB}$ denotes the location of the radiative-convective boundary or, more generally, the altitude in the planet below which the planet is well mixed, with $X(P>P_{\rm RCB}) = X_0$ and $P_{\rm RCB} > P_{\rm i}$.

\begin{figure}[t!]
\centering
\includegraphics[width=0.47\textwidth]{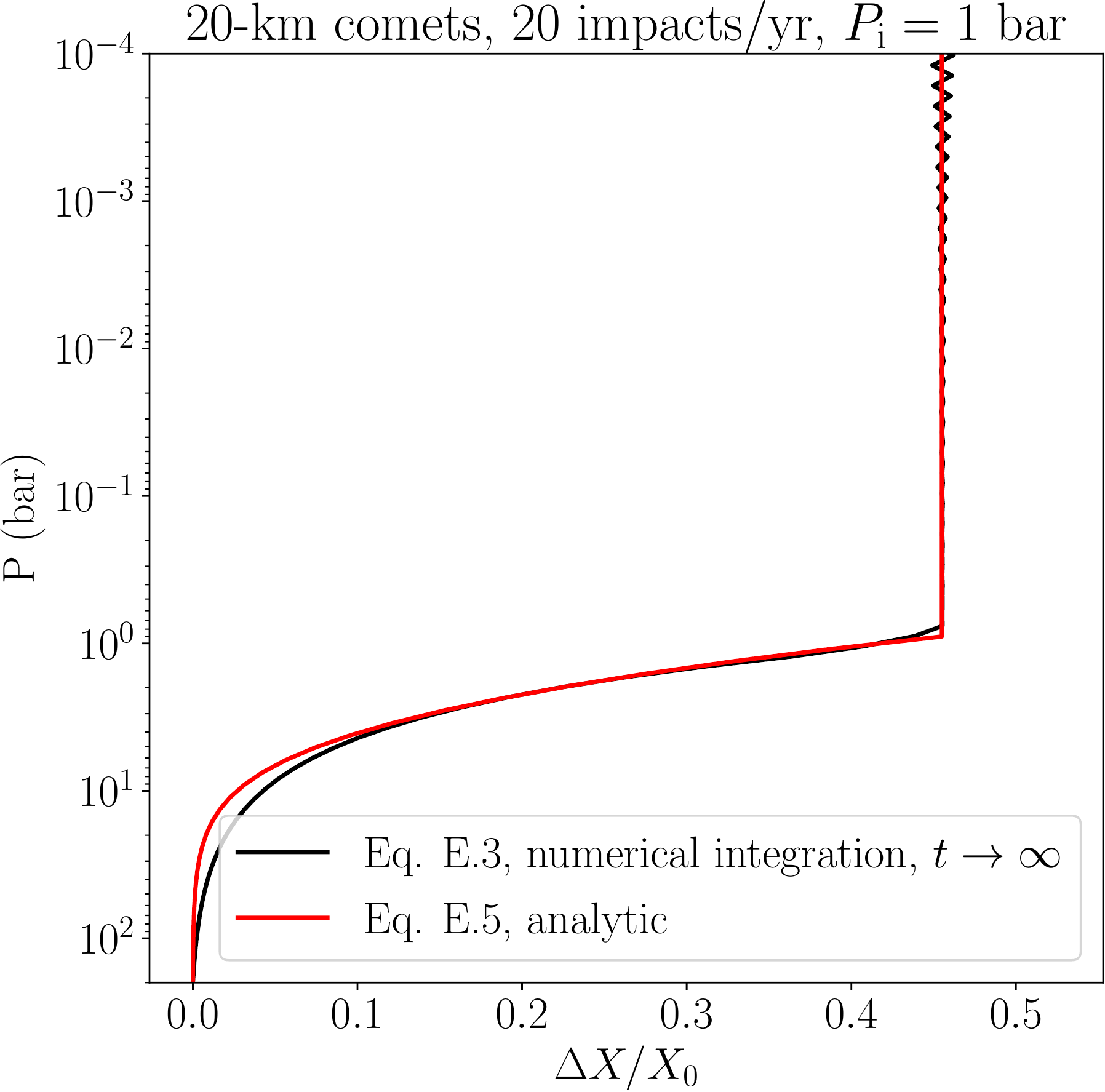}
\caption{Verification of Equation \ref{equ:deltaXApp} by comparing to the numerical integration of Equation \ref{equ:time_evo_X} to very lo ng times $t$, to obtain the limit $t \rightarrow \infty$. To obtain non-zero $\Delta X/X_0$ values, an unrealistically low value of $P_{\rm i}=1$~bar was chosen, and 20 cometary impacts per year with comets of 20~km in size were assumed.}
\label{fig:appendix_mixing}
\end{figure}

From Equation \ref{equ:deltaX} we see that $\Delta X$ will become large for small $P_{\rm i}$ and large $P_{\rm RCB}$. \citet{pinhasmadhusudhan2016} found that water ice comets of 1~km in size will be destroyed by 100 bar if impacting Jupiter at terminal velocity. From our condition that $P_{\rm RCB} > P_{\rm i}$ this means that we need to consider $P_{\rm RCB} > 100$~bar. For warm Jupiters the maximum depth of the RCB may be as deep as $P_{\rm RCB} = 200$~bar \citep{thorngrenfortney2019,sarkismordasini2021}. Then assuming $X_0=10^{-3}$, $M_{\rm P}=1 \ {\rm M_{Jup}}$, $R_{\rm P}= 1 \ {\rm R_{Jup}}$, $H_P = 200 \ {\rm km}$, $K_{zz}=10^8 \ {\rm cm^2s^{-1}}$, $P_{\rm RCB} = 200 \ {\rm bar}$ and $P_{\rm i} = 100 \ {\rm bar}$ results in a relative enrichment of $\Delta X/X_0=3\times 10^{-4}$, if a very high impact rate of $10^5$ comets of 1~km size per year are assumed. This would correspond to 275 impacts per day (and would double the total water content of the planet in $5\times 10^6$ yr). Therefore, combining reasonable estimates for the parameters describing the atmosphere with a very high cometary impact rate would not really change the water content of the planet's atmosphere (neglecting the change in $X_0$ over $5\times 10^6$ years). An obvious caveat of our toy model that comes to mind is the assumption that the planet will mix any pollution away instantaneously at pressures larger than the radiative convective boundary. The deep convective $K_{zz}$, estimated from mixing length theory, may be in the range of $K_{zz}=10^9$~cm$^2$s$^{-1}$ \citep[see, e.g., Equation 4 of][]{zahnlemarley2014}.
Extending the integration domain to $2\times 10^4$~bar and setting $K_{zz}=10^9$~cm$^2$s$^{-1}$ for $P>P_{\rm RCB}$ leads to $\Delta X/X_0=3\times 10^{-2}$ when numerically integrating Equation \ref{equ:dXdP}, that is, 100 times higher, but still  too low.

Finally, to demonstrate the good agreement of our analytical Equation \ref{equ:deltaXApp} we show a comparison to Equation \ref{equ:time_evo_X}, numerically integrated to very long times $t$, to obtain the limiting case $t \rightarrow \infty$, in Figure \ref{fig:appendix_mixing}. For this comparison 20 comets per year of 20~km size were assumed, which are destroyed at an unrealistically low $P_{\rm i}=1$~bar (to obtain non-negligible $\Delta X/X_0$ values).

\section{Toy formation models}

\subsection{\"Oberg et al. (2011) model}
\label{appendix:disk_model_Oeberg}

Here we describe the disk setup used for inverting the \citet{oeberg2011} model (see Section \ref{subsect:inverting_oeberg}), which assumes a static disk composition. More specifically, our disk model \rrp{is based on} the description in \citet{oebergwordsworth2019}, which assumes a static power-law for the disk temperature and density of the young solar nebula. Inside its iceline position a given volatile species is in the gas phase, outside it is in the solid phase. For every species the mass fraction compared to the total disk mass is tabulated, in addition to the mass fractions of the constituent atoms within a volatile species. We also account for refractory material, which we include in our framework by setting its iceline position to zero. The background gases \ce{H2} and \ce{He} are included by setting their iceline to 1000~au. This ensures that the refractory and background species stay condensed/gaseous within the simulation domain. Table \ref{tab:disk_setup} lists the mass fractions, iceline positions, and atomic composition of all considered disk species and their constituent atoms. \rrp{We note that the iceline positions given here are those expected for the disk around HR~8799, which have been obtained from our ANDES disk model, see Section \ref{subsect:formation_inversion_hr8799e}.}

\begin{table*}[t!]
\centering
{ \footnotesize
\begin{tabular}{lllll}
\hline \hline
Disk species & Constituent atoms & Relative atom mass contribution to disk species & Disk mass fraction & iceline (au) \\ \hline
Refractories &  &  & $6.2 \times 10^{-3}$ & 0$^{\rm (a)}$ \\
 & Fe & 0.21 &  &  \\
 & Mg & 0.15 &  &  \\ 
 & Si & 0.17 &  &  \\
 & O & 0.29 &  &  \\
 & C & 0.13 &  &  \\
 & P & $8.9 \times 10^{-4}$ &  &  \\
 & S & 0.05 &  &  \\
\hline
\ce{H2O} &  &  & $1.6 \times 10^{-3}$ & \rrp{2.3} \\
 & H & $2/18$ &  &  \\
 & O & $16/18$ &  &  \\ \hline
\ce{NH3} &  &  & $8.5 \times 10^{-5}$ & \rrp{6.5} \\
 & N & $14/17$ &  &  \\
 & H & $3/17$ &  &  \\ \hline
\ce{CO2} &  &  & $1.3 \times 10^{-3}$ & \rrp{8.0} \\
 & C & $12/44$ &  &  \\
 & O & $32/44$ &  &  \\ \hline
\ce{CO} &  &  & $2.3 \times 10^{-3}$ & \rrp{35.0} \\
 & C & $12/28$ &  &  \\
 & O & $16/28$ &  &  \\ \hline
\ce{N2} &  &  & $6 \times 10^{-4}$ & \rrp{73.0}\\
 & N & $1$ &  &  \\ \hline
Background &  &  & $0.987$ & 1000$^{\rm (a)}$ \\
 & H & $0.75$ &  &  \\
 & He & $0.25$ &  &  \\
\hline
\end{tabular}
}
\caption{Composition of the protoplanetary disk model and assumed iceline positions, adapting the abundances given for the young solar nebula in \citet{oebergwordsworth2019} \rrp{to the case of \rrp{HR~8799}}. Notes: (a) the icelines of the refractory and background species were set to arbitrarily small/large values to ensure that they stay condensed/gaseous within the simulation domain.}
\label{tab:disk_setup}
\end{table*}

The mass fractions were obtained using the provisional proto-solar nebula composition from \citet{oebergwordsworth2019} with some modifications, see below. The abundance of a species $i$ is given as number fractions $x_i$ in \citet{oebergwordsworth2019}, compared to the number of hydrogen atoms $n_{\rm H}$. This was converted into mass fractions $m_i$ relative to the total disk mass using
\beq
m_i = \frac{\mu_i x_i}{1.4},
\label{equ:conversion_to_mfr}
\eeq
with $\mu_i$ being the molecular mass of species $i$ in atomic mass units. This expression is obtained from the disk's total mass $M_{\rm tot}$, the mass of species $M_{i}$ of species $i$, and setting
\begin{align}
m_i & = \frac{M_i}{M_{\rm tot}} \\
& \approx \frac{\mu_i x_i n_{\rm H}}{\mu_{\rm H} n_{\rm H} + \mu_{\rm He} n_{\rm He}} \\
& = \frac{\mu_i x_i}{\mu_{\rm H} + \mu_{\rm He} n_{\rm He}/n_{\rm H}},
\end{align}
where it was assumed that most of the disk mass is contributed by \ce{H} and \ce{He} atoms. Setting $\mu_{\rm H} = 1$, $\mu_{\rm He} = 4$ as well as assuming that $n_{\rm He}/n_{\rm H} = 0.1$ \citep[see Table 8 in][for the recommended proto-solar abundances]{lodders2019} leads to the relation given in Equation \ref{equ:conversion_to_mfr}.

The refractory composition model was likewise constructed using the information given in \citet{oebergwordsworth2019}. In their model, this results in 30~\% of all oxygen in the form of refractory silicates, identical to the amount of oxygen in \ce{H2O}. We assumed that the silicates consist purely of \ce{MgSiO3}, which also conserves the solar Mg/Si abundance ratio, which is close to unity \citep{asplund2009,lodders2019}. The refractory carbon component plus volatile organics account for 50~\% of all carbon atoms, with a 3:1 ratio between the two. To simplify we added the carbon of the volatile organics by increasing the CO abundance, taking the required oxygen from the water mass reservoir. Because the fate of organic carbon, especially in the inner part of the disk, is uncertain anyway \citep[see, e.g.,][and the references within]{mordasinivanboekel2016,cridlandeistrup2019} we decided to forego a more careful treatment of the organic carbon reservoir for our conceptual study here. Iron, sulphur and phosphorous atoms were assumed to only be present in the refractory phase. The C, Fe, S, and P abundances, relative to H, were taken from \citet{asplund2009}. Our resulting C/O ratio distribution for the disk solid and gas components is shown in Figure \ref{fig:oeberg_steps}.

\subsection{Disk chemical evolution}
\label{appendix:disk_model_ANDES}

Here we describe the ANDES chemistry model used for inverting the formation model including disk chemical evolution (see Section~\ref{subsect:chemical_time_evo_inversion}), which includes the evolution of the disk's chemical composition. In ANDES, surface reactions are described by the Langmuir--Hinshelwood mechanism and are not limited to hydrogenation. H and H$_2$ tunneling through reaction barriers is also included. Any dynamical effects on the distribution of C and O, such as drifting grains, are omitted. The surface density profile is described by a power law with the exponent equal 1.5. The chemical network is based on ALCHEMIC  \citep{2010A&A...522A..42S,2011ApJS..196...25S}, with updated binding energies from \cite{2017SSRv..212....1C}. It incorporates the effects of XUV irradiation, cosmic rays, and radionuclides as ionization sources. The dust size distribution is described by a power law with $p=-3.5$ between 0.005 and 25\,$\mu$m, which reflects dust growth in disks compared to the ISM. It is used to calculate the radiation field and dust temperature in the disk's upper layers. An average grain radius of 0.35\,$\mu$m is used for calculating surface reaction rates. The disk abundances are initialized assuming that all volatile species are in the gas phase, using the same abundances as used for the static disk model, see Table~\ref{tab:disk_setup}.

\begin{figure}[t!]
\centering
\includegraphics[width=0.47\textwidth]{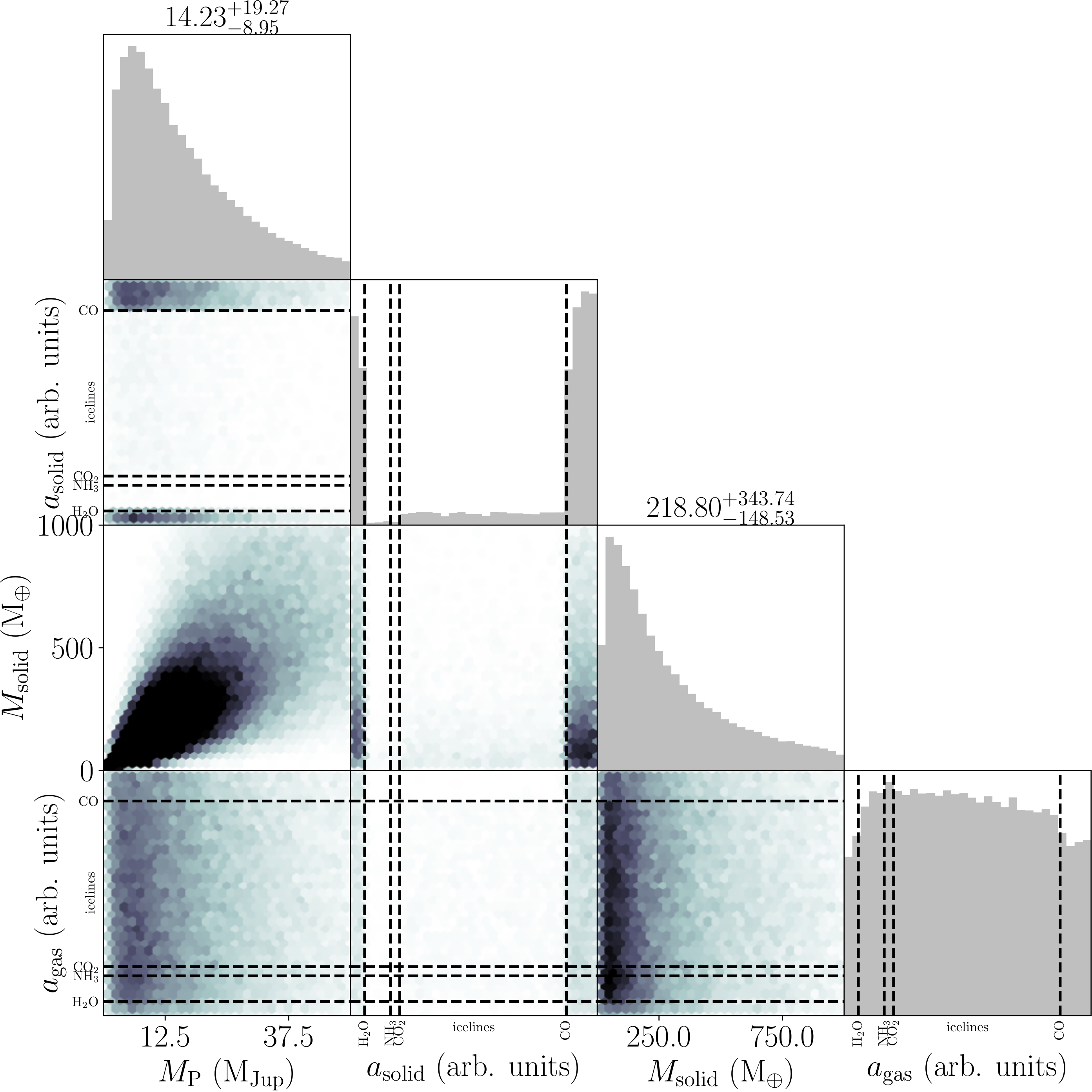}
\includegraphics[width=0.47\textwidth]{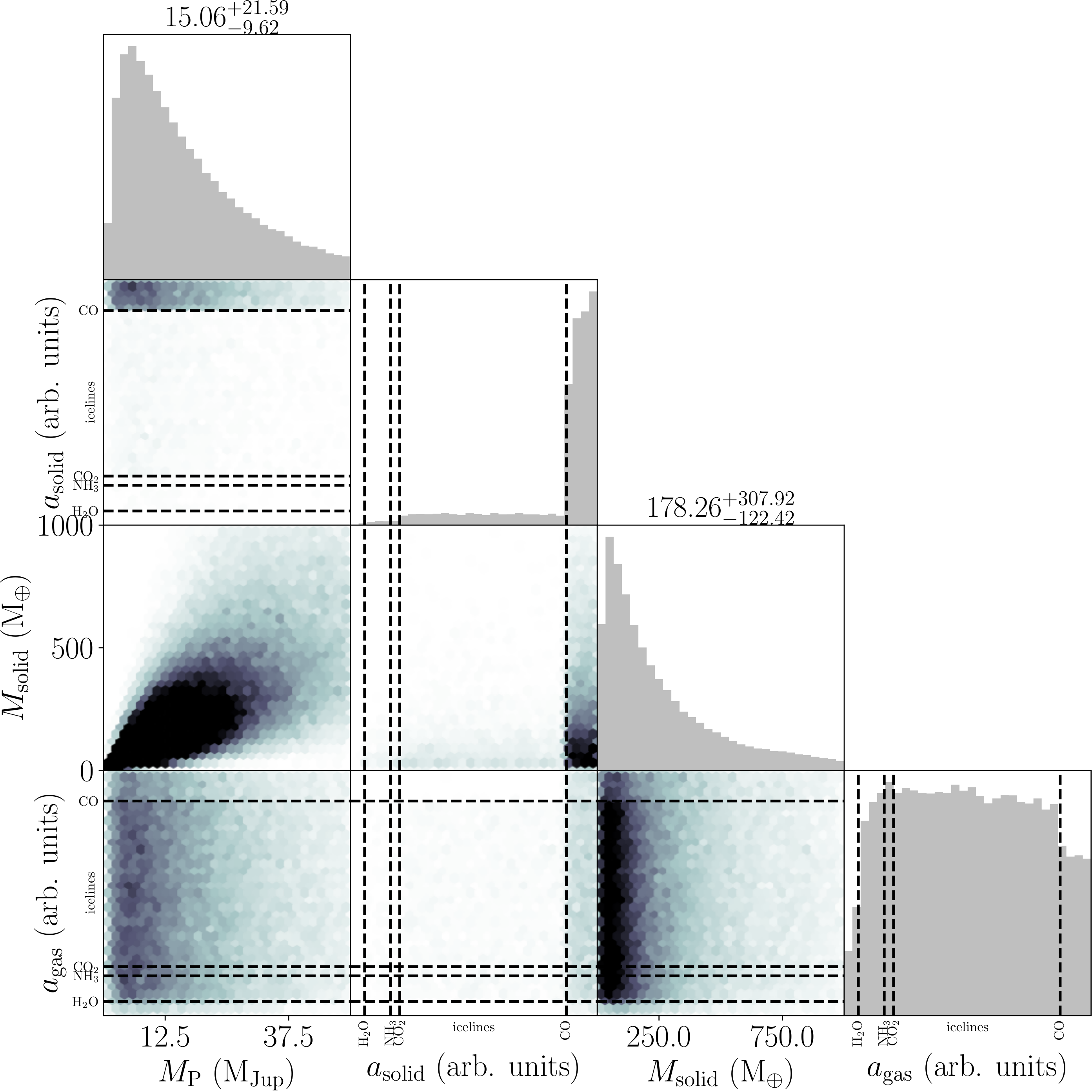}
\caption{Full posterior of the formation inversion of HR8799e with assuming the static disk model \`a la \citet{oeberg2011}. {\it Left panel:} nominal disk composition (solar). \rrp{{\it Right panel:} same as left panel, but assuming a $\lambda$~Boo-type composition for HR~8799.}}
\label{fig:hr8799_form_full_posterior}
\end{figure}

\section{Full inversion posterior of HR~8799e}
\label{app:hr8799e_full_posterior}
In \rrp{the left panel of} Figure \ref{fig:hr8799_form_full_posterior} we show the 1- and 2-d projections of the full posterior resulting from the formation inversion of HR~8799e, as discussed in Section \ref{subsect:formation_inversion_hr8799e}, using the static disk composition. The posterior of the planetary mass closely follows the mass prior, which we sampled by using the spectral retrieval results for HR~8799e, namely ${\rm log}(g)=4.0\pm 0.5$, $R_{\rm P}= 1.12\pm 0.09$~$\rm R_{Jup}$, as reported in \citet{mollierestolker2020}. We neglected the error on $R_{\rm P}$, assumed that ${\rm log}(g)$ follows a Gaussian distribution, and converted to mass via $10^{{\rm log}(g)}R_{\rm P}^2/G$, where $G$ is the gravitational constant. A flat prior was assumed on the formation/accretion locations. The prior on the accreted planetesimal mass was taken to be flat, ranging from 0 to 1000 ${\rm M}_{\oplus}$. The posterior of the accreted solid mass can be explained considering the atmospheric metallicity that was used as an input to the formation inversion (${\rm [Fe/H]}=0.48 \pm 0.25$), and the total mass of the planet. The increased probability of the planet having accreted solids from outside the CO iceline or inside the \ce{H2O} iceline (discussed in Section \ref{subsect:formation_inversion_hr8799e}) is visible. There also exists a less likely solution with lower total metallicity (lower solid mass) where both the solids and the gas were accreted within between the \ce{H2O} and CO icelines. This branch of solutions can be explained by studying Figure \ref{fig:oeberg_steps}, showing the variation in C/O in the disk gas and solids as a function of distance: within these two icelines the solids' C/O is sub-stellar and the planetary C/O can be raised to higher values by accreting gas which has an increased C/O ratio. In our current model setup this only works if the planet has a low metallicity, otherwise the gas enrichment cannot compete with the metal enrichment from the solids. \rrp{In the right panel we show the corresponding posterior in the case when $\lambda$~Boo-type abundances are assumed for the disk of HR~8799. The solution inside the \ce{H2O} iceline is no longer valid for $a_{\rm solid}$, due to the high local solid C/O ratio.}

\section{Refractory chemistry}
\label{sect:refractory_chem}

Here we give a short description of the chemical behaviour of atmospheric species that trace the planetary refractory content, as shown in Figure \ref{fig:atmo_detect_composition}. We outline their behavior as a function of temperature, assuming a pressure of 0.1 bar, whereas dissociation and ionization values were obtained from assuming pressures of 0.1 to 0.001 bar. We either assumed solar C/O ($=0.55$) or ${\rm C/O}=1.1$. We only roughly determine the transition temperatures, as these may also depend on the atmospheric gravity and metallicity. Moreover disequilibrium chemistry, internal luminosity, insolation flux and cold trapping can play important roles \citep[see, e.g.,][]{spiegel2009,fortneyvisscher2020,parmentierfortney2016}. The temperatures given here do therefore not necessarily directly translate into planetary effective temperatures. If no reference is given, we use the equilibrium chemistry code described in \citet{mollierevanboekel2016} to determine the chemical behavior.

\subsection*{\ce{H2S}}
\ce{H2S} condenses into \ce{NH4SH} at $\sim$200~K, the higher temperature condensates MnS, ZnS, \ce{Na2S} are of minor importance \citep{lodders2010}. For C/O~$<$~1, \ce{H2S} dissociates at $\sim 2000$~K, while it moves into species like CS for C/O~$>$~1 at temperatures around 1500-2000~K.

\subsection*{\ce{PH3}}
\ce{PH3} condenses into \ce{H3PO4} at 500~K. Its presence in Jupiters atmosphere at lower temperature indicates a deep quenching point however, such that \ce{PH3} may still be visible at lower temperatures \citep[see, e.g.,][]{baudinomolliere2017}. For temperatures approaching 1000~K, \ce{PH3} is increasingly converted into \ce{PH2}.

\subsection*{\ce{Na}}
Na condenses into \ce{Na2S} at $\sim 900$~K. In principle alkalis such as Na could also be sequestered into high-temperature condensates such as feldspars. However, this does likely not occur due to the rainout of silicates \citep[e.g.,][]{linemarley2017} which deplete Si from the atmosphere, which is needed for feldspar formation. Above 900~K Na is thus in the gas phase, until it gets ionized at around 2500~K.

\subsection*{\ce{K}}
K condenses into KCl at $\sim 900$~K. In analogy to Na, sequestration of K into feldspars does likely not occur due to silicate rainout. Thus, K is in the gas phase from $\sim$900~K to $\sim$2000~K, after which it is ionized. Ionization occurs at temperatures roughly 500~K cooler than for Na.

\subsection*{\ce{Refractory clouds}}
As mentioned above, the refractory cloud species \ce{Na2S} and KCl likely form at temperatures below 900~K. Here we focus on the remaining cloud species forming at intermediate to hot temperatures, and concentrate on those carrying the largest mass and/or opacity, often using the data given in \citet{wakefordvisscher2017}, or own equilibrium chemistry calculations. Refractory clouds can exist if the atmospheric temperature is below their respective evaporation temperature. Like all chemical transitions discussed here, this temperature depends on the elemental abundance and local atmospheric pressure. Under our adopted standard conditions silicates such as \ce{MgSiO3} and \ce{Mg2SiO4} evaporate at temperatures around $\sim$1600~K, while iron clouds are stable until $\sim$1700~K. VO and calcium titanates are stable until 1600 to 1800~K, respectively. Aluminum-bearing condensates such as \ce{Al2O3}, which are among the most stable ones, evaporate around 1900~K. Among the species listed here, potentially only silicates, \ce{Al2O3}, and KCl may actually form in the visible part of the atmosphere, as these species have low surface energies, leading to high nucleation rates \citep{gaothorngren2020}.
The cloud bases will reside deeper inside the planetary atmosphere for lower temperatures, with the cloud particles entering from above, or settling below the photosphere. For brown dwarf this temperature-dependent removal of silicate clouds is thought to cause the L-T transition, which typically occurs at $T_{\rm eff}=1200$ to $1400$~K \citep[e.g.,][and the references therein]{bestliu2021}, while for planets and low-gravity brown dwarfs this limiting temperature may be as low as approximately 1000~K \citep[e.g.,][]{morleyfortney2012,marleysaumon2012,charnaybezard2018}.

\subsection*{SiO}
SiO is an especially interesting molecule for tracing the abundance of the refractory silicates in the atmosphere such as \ce{MgSiO3} or \ce{Mg2SiO4}. As soon as the silicates evaporate (around 1600~K) their constituent atoms move into the gas phase. While atomic Mg is then the preferred gaseous form of Mg, Si will move into SiO \citep{visscherlodders2010}. For C/O~$\gtrsim 1$, SiO enters the gas phase at around 1300~K, which is when SiC evaporates. SiO then starts to dissociate around 3500~K for C/O~$\lesssim 1$, while moving into SiS around 2000~K for C/O~$\gtrsim 1$. The local evaporation-dependent temperatures given here could be lower than the observed transition as a function of planetary effective temperature, where high-pressure cloud formation could cold-trap Si into silicates.

\subsection*{Metal oxides}
Similar to SiO the other metal oxides form as soon as the refractory clouds evaporate. Possible species of interest are TiO, VO, SiO, AlO, CaO \citep[e.g.,][]{sharpburrows2007,gandhimadhusudhan2019}, with the relevant evaporation temperatures of the clouds ranging from $\sim$1600-1900~K at our adopted standard condidtions. Again, these are then only expected to be visible in the atmosphere if not cold-trapped into condensates at lower altitudes, that is, higher pressures. Except for SiO (see discussion in the SiO section above) most of these metal oxides are not expected to form in atmospheres with C/O~$\gtrsim$~1 \citep{madhusudhan2012,gandhimadhusudhan2019}. In general, metal oxides will be destroyed by dissociation at high enough temperatures, with TiO and VO dissociating at temperatures similar to water (around 3000~K). As stated above SiO is a bit more stable, dissociating from temperatures higher by about 500~K.

\subsection*{Metal hydrides}
Similar to metal oxides, metal hydrides such as FeH, CaH, MgH, NaH, CrH, TiH may form as soon as the refractory clouds have been evaporated, thus at local atmospheric temperatures of around 1600 to 1900 K and cooler temperatures for NaH, as \ce{Na2S} evaporates at $\sim$900~K already. Of course the cold trapping statement from above holds here as well. The hydrides will be destroyed by thermal dissociation at high enough temperatures, for example MgH and FeH dissociate around 3000~K or so \citep{lothringerbarman2018}. For these two species it should also be noted that the main gas phase carrier is atomic Mg and Fe in (ultra) hot Jupiter atmospheres, and that MgH and FeH are less abundant by about a factor $10^4$ \citep{visscherlodders2010}.

\subsection*{Metal atoms}
Metal atoms can exist in the gas phase as soon as the sequestering refractory condensates evaporate (modulo cold trapping). As mentioned above, gaseous Fe and Mg are the main gas carriers of these elements once the silicates and iron condensates are gone. Si takes over as the main gas species only after SiO is dissociated. Fe, Mg, Ti, Ca, Ni all are ionized between 3500 to 4000 K or so, with Al being ionized at somewhat lower temperatures \citep{lothringerbarman2018,hoeijmakersehrenreich2019,kitzmannheng2018}.

\subsection*{Metal ions}
Metal ions become visible in the atmosphere as soon as the atoms have been ionized, see immediately above.

\bibliography{composition_formation}{}
\bibliographystyle{aasjournal}



\end{document}